\documentclass[journal]{IEEEtran}
\pdfoutput=1

\usepackage{cite}
\usepackage{amsmath,amssymb,amsfonts}
\usepackage{algorithm}
\usepackage{graphicx}
\usepackage{textcomp}
\usepackage{graphics} 
\usepackage{epsfig} 
\usepackage{mathptmx} 
\usepackage{times} 
\usepackage{mathrsfs}
\usepackage{mathtools}
\usepackage{stfloats}
\usepackage[noend]{algpseudocode}
\usepackage{multirow}
\usepackage{subcaption}
\usepackage{url}

\usepackage{amsthm}
\theoremstyle{definition}
\newtheorem{definition}{Definition}[section]

\newsavebox{\bigimage}

\def\BibTeX{{\rm B\kern-.05em{\sc i\kern-.025em b}\kern-.08em
    T\kern-.1667em\lower.7ex\hbox{E}\kern-.125emX}}

\begin{document}

\title{Adaptive Workload Allocation for Multi-human Multi-robot Teams for Independent and Homogeneous Tasks}

\author{Tamzidul Mina$^{1}$, Shyam Sundar Kannan$^{2}$, Wonse Jo$^{2}$, and Byung-Cheol Min$^{2}$

\thanks{$^{1}$Tamzidul Mina is with the SMART Lab, Department of Computer and Information Technology, Purdue University, and with the School of Mechanical Engineering, Purdue University, 
		West Lafayette, IN 47907, USA
        {\tt\small tmina@purdue.edu}}%
\thanks{$^{2}$Shyam Sundar Kannan, Wonse Jo, and Byung-Cheol Min are with the SMART Lab, Department of Computer and Information Technology, Purdue University,
        West Lafayette, IN 47907, USA
        {\tt\small \{kannan9,jow,minb\}@purdue.edu}}%
\thanks{This work was supported in part by NSF CAREER Award IIS-1846221.}
}

%



\maketitle

\begin{abstract}
Multi-human multi-robot (MH-MR) systems have the ability to combine the potential advantages of robotic systems with those of having humans in the loop. Robotic systems contribute precision performance and long operation on repetitive tasks without tiring, while humans in the loop improve situational awareness and enhance decision-making abilities. A system's ability to adapt allocated workload to changing conditions and the performance of each individual (human and robot) during the mission is vital to maintaining overall system performance. Previous works from literature including market-based and optimization approaches have attempted to address the task/workload allocation problem with focus on maximizing the system output without regarding individual agent conditions, lacking in real-time processing and have mostly focused exclusively on multi-robot systems. Given the variety of possible combination of teams (autonomous robots and human-operated robots: any number of human operators operating any number of robots at a time) and the operational scale of MH-MR systems, development of a generalized framework of workload allocation has been a particularly challenging task. In this paper, we present such a framework for independent homogeneous missions, capable of adaptively allocating the system workload in relation to health conditions and work performances of human-operated and autonomous robots in real-time. The framework consists of removable modular function blocks ensuring its applicability to different MH-MR scenarios. A new workload transition function block ensures smooth transition without the workload change having adverse effects on individual agents. The effectiveness and scalability of the system's workload adaptability is validated by experiments applying the proposed framework in a MH-MR patrolling scenario with changing human and robot condition, and failing robots. 
\end{abstract}


%
\IEEEpeerreviewmaketitle

\section{Introduction}
\IEEEPARstart{M}{ulti-human multi-robot}  (MH-MR) systems have an immense potential for applicability in various independent and non-sequential tasks such as coverage problems of surveillance, patrolling, search and rescue, inspection or assembly of items in an industrial conveyor belt by robotic manipulators, and various other multi-agent scenarios. Robots allow long operation hours on repetitive tasks and provide consistent and precise performance beyond human capability, while human operators contribute improved situational awareness, experienced and intuitive decision making, and the ability to work around unexpected situations. While research on human-robot interaction has gained a lot of momentum in recent years \cite{khoramshahi2019dynamical,villani2018survey,shiomi2018should}, MH-MR systems are a relatively new area involving interaction and collaboration between multiple humans and robots. 
\begin{figure}[t!]
    \centering
    \includegraphics[width=\linewidth]{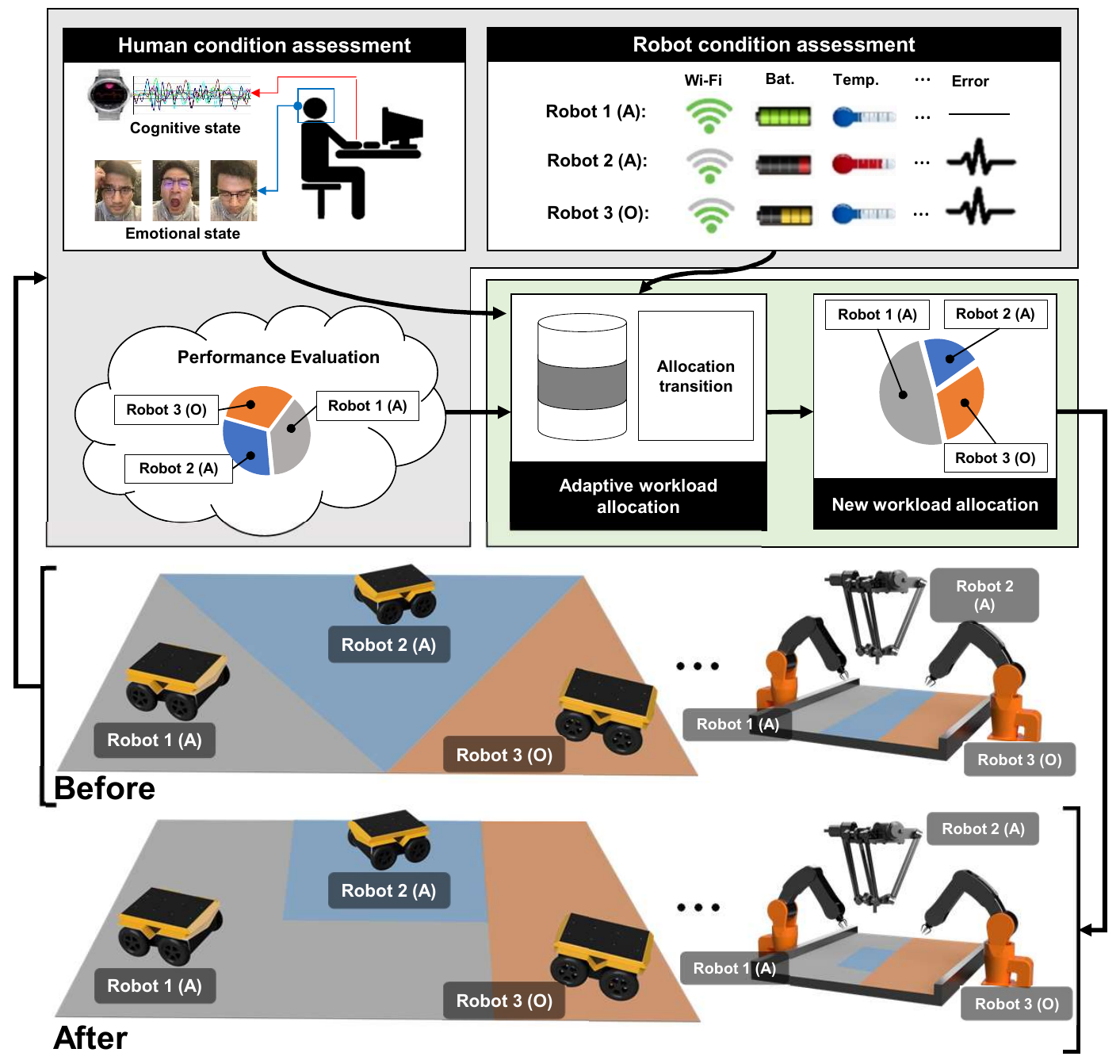}
    \caption{Conceptual illustration of the proposed multi-human multi-mobile-robot (MH-MR) system with adaptive workload allocation to human and robot conditions and performance with workload transitional considerations. Potential application includes autonomous and multi-human operated multi-mobile robot patrolling, surveillance, multi-robot manipulator tasks on a moving conveyor belt in an industrial setting etc. The dynamically allocated \textit{workspace} in the different applications change with human and robot operator condition and performance in real-time.}
    \vspace{-5mm}
    \label{fig:concept}
\end{figure}

Task/workload allocation is an important problem in MH-MR systems. Previous works have investigated team organization \cite{lewis2010teams}, a number of operator-mediated robot control methods \cite{chien2012scheduling}, awareness studies in human-robot systems \cite{drury2003awareness}, and various classifications of human-robot systems \cite{yanco2004classifying} for task/workload allocation. Tsarouchi \textit{et al.} introduced a system for designing and assigning tasks to operators and human workplaces \cite{tsarouchi2017human}. Automation adaptation based on human perceived workload has also been studied in \cite{kaber2006situation}. Physiological measurements of humans have been used as triggers in the control of unmanned aerial vehicles (UAVs) to initiate different workload states and adapt operator performance \cite{prinzel2000closed,wilson2003operator}. Error rates and task difficulty as perceived by operators have also been used as triggers to re-allocate or automate workload \cite{parasuraman2007adaptive}. However, task allocation in a multi-agent system increases in complexity if the triggers are less than perfect; sudden or unpredictable changes in workload or mission may have a negative impact on the operator's performance pertaining to the understanding of the automation behaviors and the system functions they control \cite{miller2007designing}. Sudden and/or drastic changes may overwhelm or momentarily catch human operators off guard while trying to cope with their allocated work.  

Musi\'{c} and Hirche have proposed an architecture for planning human roles in robot team control \cite{music2017control} to optimize collaboration and teaming mechanisms across a wide range of human operators and robots. Task allocation in multi-robot scenarios have been widely studied in \cite{lee2018resource,setter2016energy,koubaa2018general,booth2016constraint,otte2020auctions}, considering resource constraints and robot performance. Task allocation with unknown robot capabilities have also been studied in \cite{emam2020adaptive}. Optimal task allocation with multi-humans in the loop has also been proposed in \cite{malvankar2015optimal}, where task allocation is performed over multiple-levels (group and individual) comprising of high-risk and low-risk information in order to maximize effectiveness of the entire system minimizing processing cost and time, considering human factors given limited resources. In market-based approaches for multi-agent task allocation, the team seeks to optimize an objective function based upon robots utilities for performing particular tasks \cite{tang2007complete,zlot2006market}; desirable features of these approaches include efficiency in satisfying the objective function, robustness and scalability of the system. However, in systems where fully centralized approaches are feasible, market-based approaches can be more complex to implement and can produce poorer solutions; when fully distributed approaches suffice, market-approaches can be unnecessarily complex in design and can require excessive communication and computation \cite{dias2006market}. Mixed integer linear programming optimization approaches have also been used for task allocation \cite{atay2006mixed,darrah2005multiple,mosteo2006simulated}. Population based approaches such as the genetic algorithm was also proposed for task allocation in disaster scenarios \cite{jones2011time}. Ant colony optimization has also been proposed for task allocation of multi-agent systems in \cite{wang2012multi}. 

Most of the task/workload allocation methods proposed in literature have focused on maximizing the system's work output without considering individual agent conditions. Moreover, most of the research work on task allocation has remained confined to multi-robot systems only. In contrast, we present a task/workload allocation method considering quantified both human and robot condition and performance equally, i.e. prioritizing the \textit{ability} of all agents to work in an MH-MR system maintaining agent level work efficiency, while ensuring full coverage of the application workspace.

In this paper, we present a generalized MH-MR framework capable of workload allocation for independent, non-sequential homogeneous tasks, consisting of independent modular function blocks assessing human and robot conditions and the performances of human-operated and autonomous robots. The system is designed to be compatible with previously established normalized quantitative human and/or robot health and performance assessment tests. The framework also incorporates a re-allocated workload transition model to minimize the effects of sudden changes in workload or mission that may have negative impacts on operator or robot performance. We demonstrate the applicability, effectiveness, and scalability of the framework through various scenarios of a MH-MR patrolling application as validation of the proposed concept. An overview of our MH-MR work allocation concept applied in example scenarios of mobile robot coverage problems and robotic manipulators in an assembly line conveyor belt is shown in Fig. \ref{fig:concept}.

\section{Preliminaries and Assumptions}\label{sec:prob_stat}
We consider a homogeneous group of $m$ robots capable of carrying out autonomous missions, each denoted as $R_i$, for $i\in I_R=\{1,2,..,m\}$ with state definition of $\mathbf{q}_i \in \mathbb{R}^w$, where $w$ represents the dimension of the system workspace. $R_i$ may be teleoperated by any number of human operators at any time, each denoted as $O_j$ for $j\in I_O=\{1,2,..,h\}$, modeled as an edge $E$ in an undirected graph $\mathscr{G}=(V,E)$ without any self-connectivity, where $V$ represents the set of nodes $(R_i,O_j), i \in I_R, j \in I_O$, such that $\{R_i, O_j\} \in E$. We denote the set of indices of human-operated robots as $I_H$ and the set of indices for autonomous robots in the system as $I_A$. For the convenience of the reader, we summarize the terminology usage in this section as: $r/R$ for robots, $o/O$ for human Operators, $c/C$ and $p/P$ for condition and performance (human and robot).

Each human operator may control multiple robots and assume/relinquish control of any robot in the system at any time, triggering a change in the robot operation mode. We do not limit robots operated by humans to be only teleoperated; some level of autonomy might exist while the human operator acts as a supervisor. Regardless, the performance of such a robot is dependent on the state of the human operator as well.

The condition or \textit{health status} of each robot $R_i$, for $i\in I_R$ in the MH-MR system can be monitored at all times as a set of robot \textit{health} states denoted as $C^{R_i} \in \mathbb{R}^{w_r}$, where $w_r$ equals the number of robots in the system $m$. Physiological measurements and the emotional state of each human operator $O_j$, for $j\in I_O$ in the MH-MR system can be monitored at all times as a set of human operator \textit{health} states denoted as $C^{O_j} \in \mathbb{R}^{w_o}$, where $w_o$ is the number of human operators in the system $h$.

\begin{definition}The performance of each robot (autonomous and human-operated) $R_i, i \in I_R$ on their respective allotted mission/task can be evaluated based on a predefined evaluation metric relevant to the mission/task using observation set $P^{R_i} \in \mathbb{R}^{w_r}$, $i \in I_R$.
\end{definition}
\begin{figure*}[t!]
    \centering
        \includegraphics[width=\linewidth]{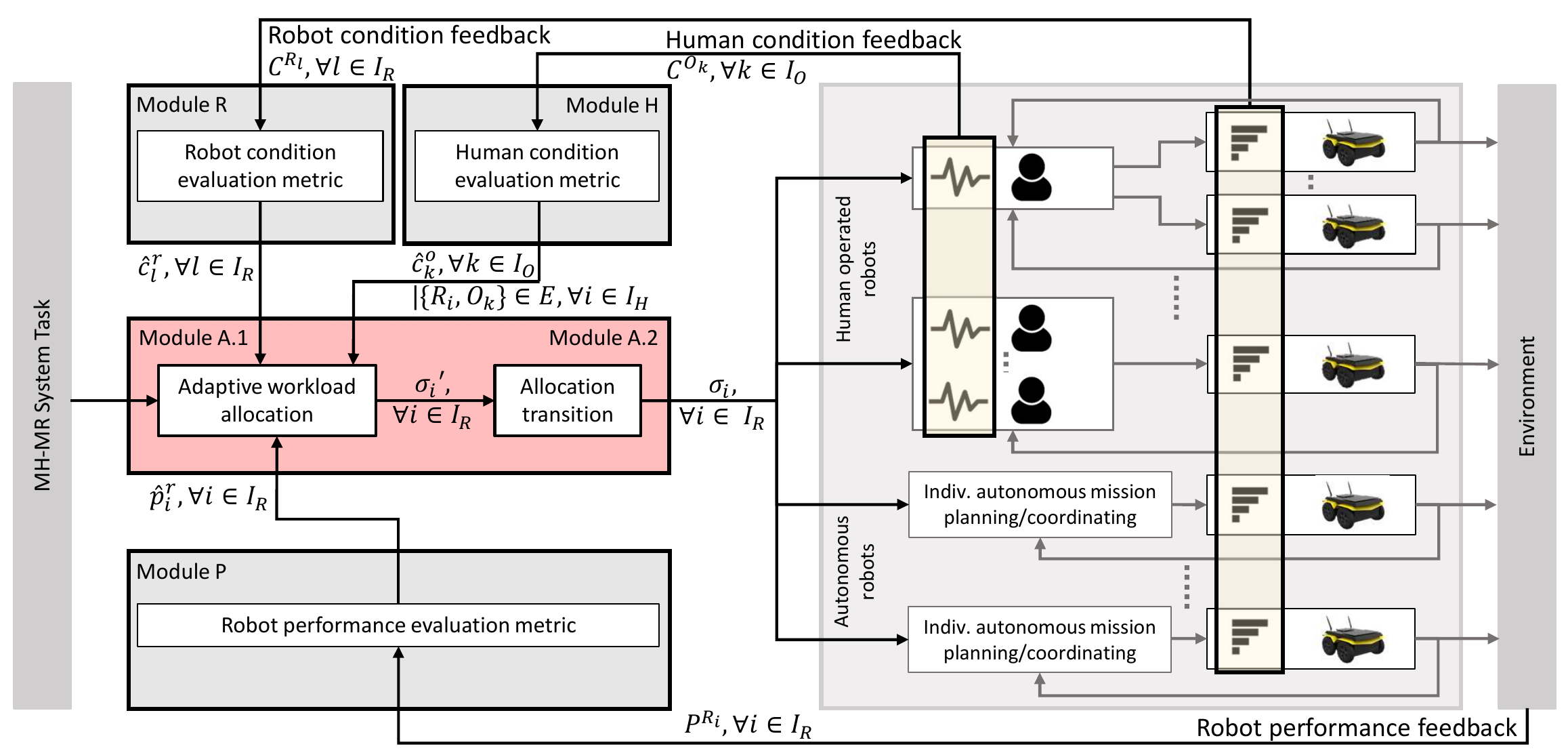}
        \caption{Adaptive MH-MR workload allocation system modular framework based on individual human and robot condition and performance. The MH-MR system may consists of autonomous robots and combinations of single human-operated single robot, multi-human operated single robots and/or single human operated multi-robots. The workload allocation module takes equal weighted metric inputs from each modular condition and performance evaluation module to allocate new workload. The allocation transition module ensures a smooth transition to the new workload.}
        \vspace{-5mm}
    \label{fig:framework}
\end{figure*}

We define the constraints of the system workspace (workload) $\mathcal{W} \in \mathbb{R}^w$ as finite, known apriori, and covered by $m$ robots without any overlap. We assume each robot is equipped with appropriate low level velocity/position controllers with collision avoidance relevant to the MH-MR application and is capable of fully autonomous behavior when required \cite{fox1998hybrid, pandey2017mobile,martinez1994fuzzy}. Once a mission is assigned to a specific robot, an autonomous robot uses its own individual mission planning/coordinating algorithms to conduct the mission. Individual mission/workload assigned to human-operated robots is coordinated by their human counterparts. 

\begin{definition}We define workload on $R_i, i \in I_R$ (either autonomous or human-operated) at time $t$ as $\sigma_i(t)$ and the corresponding workspace as $\mathcal{W}_i \in \mathbb{R}^w$ regardless of its task depending on the application. Upon mission assignment to the MH-MR system, the initial workload for each robot $\sigma_i(0)$ may or may not be equally distributed.
\end{definition}

The objective is to provide a systematic approach to an adaptive workload allocation in MH-MR systems based on: (a) robot and human operator state monitoring ($C^{R_i}, i\in I_R$ and $C^{O_j}, j\in I_O$), and (b) autonomous robot and human-operated robot work performance states ($P^{R_i}, i\in I_R$). The transition process between workload changes for a robot or operator must consider the effect of the prescribed change. The framework must maintain generality for applicability in any MH-MR system.

At the core of the proposed adaptive MH-MR system framework, the adaptive workload allocation system and a workload transition system, designated as module A.1 and module A.2 respectively, provide a workload distribution solution on the assigned mission based on the condition and performance of each human and robot unit operating in the system. The state of each human and robot in the system, and the performance of each autonomous and human-operated robot are assessed to adaptively re-allocate the total mission workload for continuous performance. 

We realize that the relevance of such evaluations or assessments are application specific and may be irrelevant in certain systems. Therefore, to maintain generality of our adaptive MH-MR system framework, we propose a modular design consisting of robot and human state and performance assessment function blocks. Each module provides a real-time metric of its unit system adhering to its own procedure based upper and lower bounds. The modules and their metrics are listed as follows:
\begin{itemize}
    \item Module R: Robot state monitoring and evaluation metric $c^r_l \in [u_R, v_R]$ from $C^{R_l}, l\in I_R$, normalized as $\hat{c}^r_l$
    \item Module H: Human operator state monitoring and evaluation metric $c^o_k \in [u_O, v_O]$ from $C^{O_k}, k\in I_O$, normalized as $\hat{c}^o_k$
    \item Module P: Robot (human-operated or autonomous) performance assessment metric $p^r_i \in [u_P,v_P]$ from $P^{R_i}, i \in I_R$, normalized as $\hat{p}^r_i$,
\end{itemize}
where $u$ and $v$ represent the lower and upper bounds of the corresponding metric respectively.

The aforementioned human and robot states and performance metrics from the modular function blocks are fed into the adaptive workload allocation module A.1 for workload re-allocation. The workload is re-allocated to maximize overall system performance at all times. The modular design ensures that any module may be added or removed from the system depending on the application requirement, pertaining to the generalization of the framework. Fig. \ref{fig:framework} illustrates the proposed modules of the MH-MR system framework. 

\section{Adaptive MH-MR System Framework}\label{sec:prob_sol}
\subsection{Module A.1 Adaptive Workload Allocation}\label{sec:work_alloc}
We design the workload allocation of the system based on the maximum outcome of combining the incoming human and robot states and performance metrics. A variant of the softmax function, also known as the normalized exponential function is proposed to determine the workload allocation for each of the $m$ robots. We define a vector \textbf{S} of $m$ normalized inputs such that,
\begin{equation}\label{eq:S_set}
    [S]_i = \begin{cases} \frac{\gamma_i}{|\Lambda_i|+2}\left(\hat{c}^r_i + \sum_{k \in \Lambda_i}\hat{c}^o_k + \hat{p}^{r}_i\right) &\mbox{if } i \in I_H \\
                        \frac{\gamma_i}{2}\left(\hat{c}^r_i + \hat{p}^{r}_i\right) &\mbox{if } i \in I_A 
                        \end{cases}
\end{equation}
where $\Lambda_i$ is a vector of $\lambda \in I_O | \{R_i,O_{\lambda}\} \in E$, $\gamma_i = \text{min}(\hat{c}^r_i, \hat{c}^o_k,\hat{p}^{r}_i)$, $\forall k \in \Lambda_i$, and  $\gamma_i =\text{min}(\hat{c}^r_i,\hat{p}^{r}_i)$.
By this design, the $\gamma$ terms ensure that the system allocates zero workload to a robot (autonomous or human-operated), if the corresponding robot and/or human operator is detected to have voluntarily/involuntarily stopped working ($\hat{p}^r$ would equal to zero), completely failed, incapacitated and/or may have suffered from any discontinuity or disconnectedness in the teleoperation and communication graph structure ($\hat{c}^r$ and/or $\hat{c}^o$ would equal to zero). It also ensures that the workload allocated is proportional to the worst human/robot condition/performance in situations where one increases and another decreases equally.

We calculate the share of the total workload for robot $R_i$, $i \in I_R$ at current time $t$ as,
\begin{equation}
    \mathbf{\sigma}_i'(t) = \frac{\mathbf{S}(i)}{\sum_{l=1}^m \mathbf{S}(l)} \quad \text{for $i={1,...,m}$}.
\end{equation}
The normalization ensures that the sum of all $\sigma'$ is $1$, pertaining to the total workload of the system.

\subsection{Module A.2 Workload Allocation Transition}
Sudden changes in workload allocation may have overwhelming effects on a host. The transitions must be smooth and manageable without any drastic changes. We model such a transition process for the workload change from the current actual workload allocation $\sigma_i(t), i \in I_R$ to the proposed workload allocation $\sigma_i'(t), i \in I_R$ considering the effect of the change to the highest affected agent in the system as follows. 

We model the workload transition process as,
\begin{align}\label{eq:sigma}
    \sigma_i(t+1) = \sigma_i(t) + K_{e}\Delta \sigma_i(t)
\end{align}
where $\Delta \sigma_i(t) = \sigma_i'(t)-\sigma_i(t)$, and $K_e \in \left[0,1\right]$ is a transition model coefficient dependent on the highest effect of the proposed change on the system.  

Denoting the proposed 2-D workspace for $R_i, i \in I_R$ corresponding to proposed workload $\sigma_i'(t)$ as $\mathcal{W}_i'$, we determine the highest effect on the system as $q_{f} = min(q_c)$, where $q_c$ denotes the shortest Euclidean distance between the boundary of workspace $\mathcal{W}_i'(t)$ and $q_i(t)$, $\forall i \in I_R$. In situations where a complete robot failure occurs or a human operator is incapacitated, the failed robot is ignored in $q_f$ determination.

The transition coefficient $K_e$ can therefore be modeled as an exponential function of $q_{f}$,
\begin{equation}\label{eq:Ke}
    K_{e} = 1-e^{-Kq_{f}}
\end{equation}
where $K$ is a positive scaling constant. The exponential nature of the transition allows for a smooth change in workload where $K$ may be tuned to control the rate of transition depending on particular application scenarios. 

The workload allocation cycle must be synced with the contributing modules in the framework. The workload allocation update cycle time constant can therefore be set as $\tau = \text{max}\left(\tau_{p^r},\tau_{c^o},\tau_{c^r}\right)$ where $\tau_{p^r}$, $\tau_{c^o}$ and $\tau_{c^r}$ denote the required operation cycle time constants for function modules P, H, and R respectively.   

\subsection{Human and Robot, Condition and Performance}
\subsubsection{Module H: Human Condition Evaluation Metric}
We define human operator condition as their ability to perform their task of teleoperating robots as a function of stress, emotion, and/or direct physiological measurements depending on the MH-MR application. For human operator condition evaluation, we refer to previous studies in literature for quantitative and qualitative techniques. Primary approaches include predicting stress or emotion from audio signals \cite{lu2012stresssense}, gestures \cite{noroozi2018survey,garcia2015automatic}, facial expressions \cite{lajevardi2012facial}, body gestures \cite{giakoumis2012using} or physiological signals such as heart rate, skin conductance, and respiration \cite{zhao2016emotion,carneiro2012multimodal,villani2017natural,sano2013stress,sun2010activity,muaremi2013towards}. The measurements and predictions are used to 
evaluate stress and psychological dynamics in the interest of creating effective working conditions \cite{greene2016survey}. Individual or a combination of a number of emotional responses may be measured and used as human operator condition for Module H, but at this stage we focus on human operator stress levels that have been shown to have a direct negative impact on work performance \cite{westman1996inverted}.  

Galvanic skin response (GSR) or skin conductance is a reliable indicator of stress \cite{healey2005detecting}. Under stress, skin conductance of an individual is increased \cite{liao2005real} due to increase in moisture on the surface of the skin, which increases the flow of electricity. 
Healey \textit{et al.} proposed a continuous stress measurement metric in \cite{healey2005detecting} that can be normalized and used as a measure of the human operator condition directly for our proposed framework. 
Also, a number of other such human operator condition measurement metrics based on facial expressions, body gestures, heart rate and respiration have been summarized in the stress recognition literature survey \cite{sharma2012objective} that may be used as Module H in our proposed framework.

Stress detection using a combination of multiple noninvasive physiological variables such as galvanic skin responses, blood volume pulse, pupil diameter and skin temperature have been proposed in \cite{zhai2006stress}. A support vector machine is used to perform the supervised classification of effective states between "stress" and "relaxed". Stress levels may also be further discretized as "low", "medium" and "high"; such discrete states may be quantified as discrete human operator condition values simply as $0.75$, $0.5$ and $0.25$ respectively, or a moving average may also be applied depending on the application. We stress here that our proposed method is designed for continuous human operator condition values, but may still be adapted with a discrete human operator condition evaluation system as well appropriate of the application.

\begin{figure*}
    \begin{subfigure}{0.48\textwidth}
    \centering
        \includegraphics[width=1\linewidth]{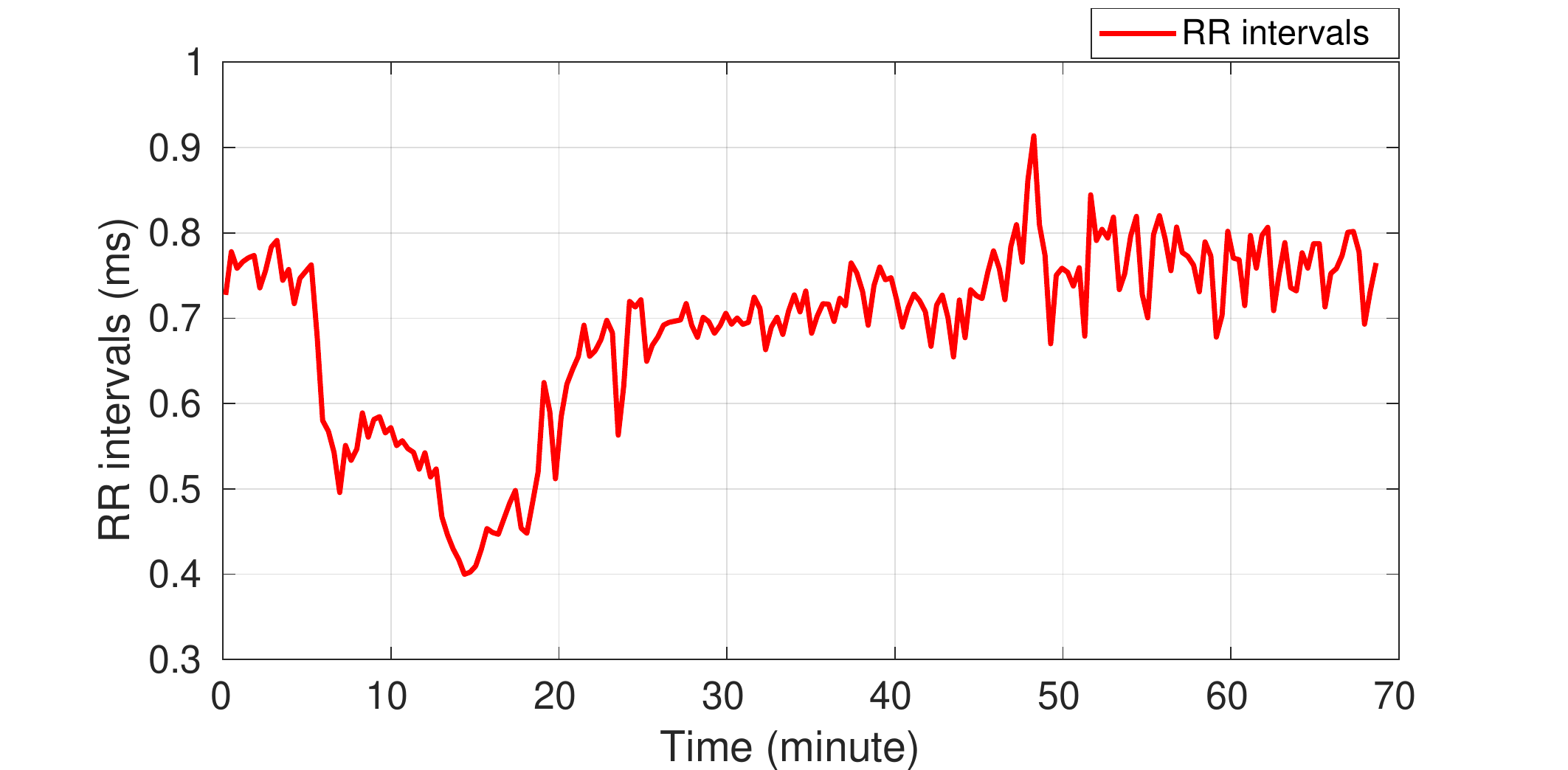}
        \caption{Raw data of R-R intervals from \cite{healey2000wearable}. R-R interval was measured from participants for a walking task between $5 - 20$\,mins followed by a small recovery period, and then a task to watch a horror film between $30-70$\, mins.}
        \label{fig:raw_bio_sensor_data}
    \end{subfigure}
    ~
    \begin{subfigure}{0.48\textwidth} 
    \centering
        \includegraphics[width=1\linewidth]{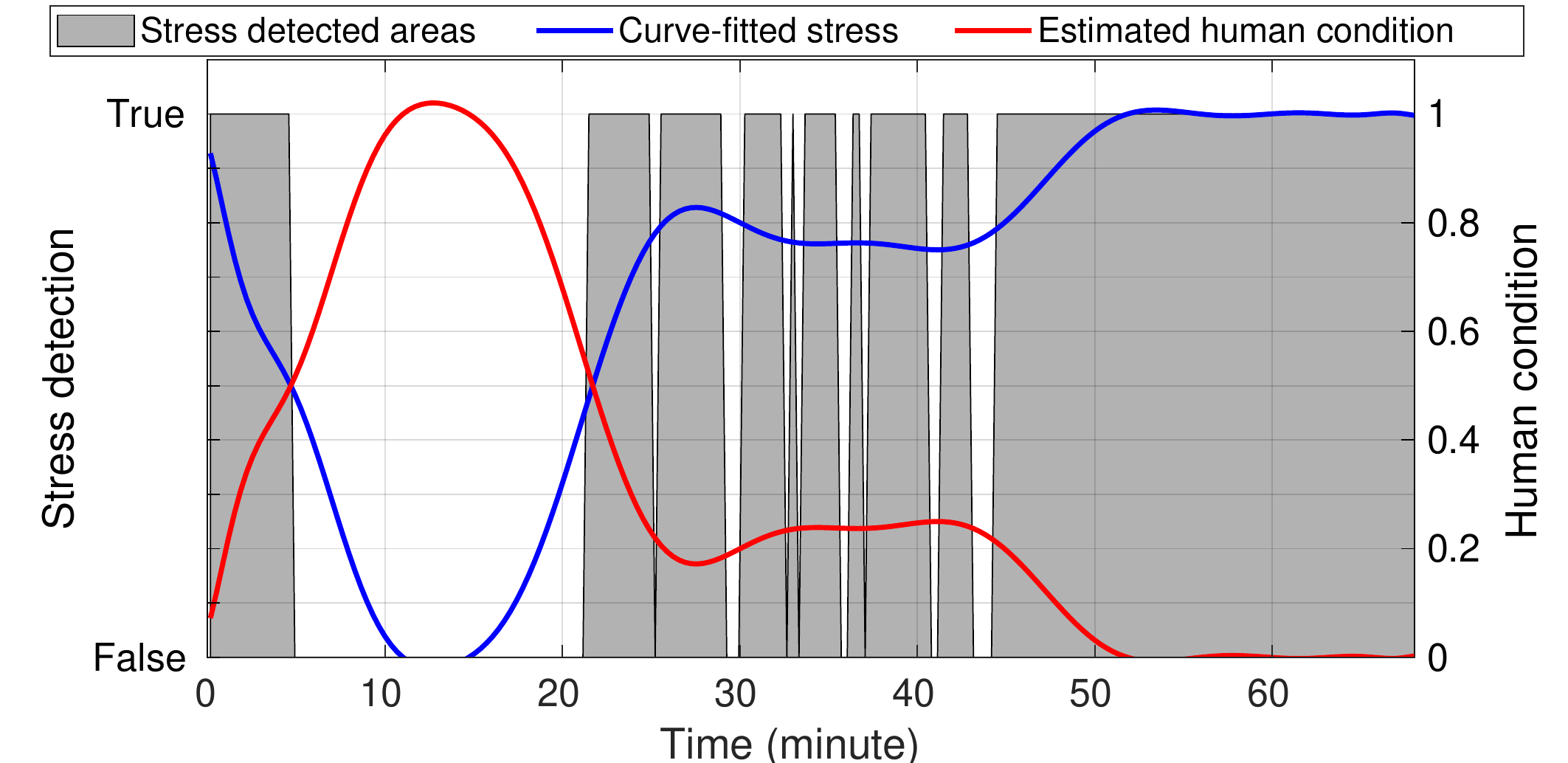}
        \caption{Stress detection using the TPOT algorithm \cite{le2020scaling}. A possible moving average filter implementation for continuous stress quantification (blue) is shown followed by its human condition metric assessment (red).}
        \label{fig:stress_detection_data}
    \end{subfigure}
    \caption{An example implementation of a machine learning-based stress detection algorithm using HRV signals and R-R intervals; (a) raw data of R-R intervals from the existing dataset \cite{healey2000wearable}, and (b) the stress detection and subsequent human condition assessment.}
    \label{fig:stress_detection_ML}
    \vspace{-5mm}
\end{figure*}

Heart rate variability (HRV) in terms of the length between heartbeat intervals, also called an R-R interval or inter-beat interval (IBI) plays an important role in predicting human condition in neurosciences and medical fields \cite{kim2018stress}. HRV has been utilized to predict stress in various works of literature \cite{munla2015driver,mcduff2014remote,gouin2015high,jovanov2003stress}, due to its responses to physiological and environmental stimulus. Ottesen proposed a stress detection algorithm using wearable devices and machine learning technology \cite{mastersthesis} using both heart rate (HR) and HRV as a training dataset from \cite{healey2000wearable} as shown in Fig. \ref{fig:raw_bio_sensor_data}; a machine learning model was proposed for automated machine learning and the evolutionary algorithm called TPOT \cite{le2020scaling}. The model had a stress detection accuracy of $79.9$\% using an existing dataset from a user study, where participants watched a horror movie after a 15-minute walking task to differentiate between physical and mental stress, such as lowering the R-R intervals. An implementation of the TPOT algorithm for stress detection is shown in Fig. \ref{fig:stress_detection_data}. 

An example implementation of the moving average filter as a possible continuous stress quantification method is shown on this discrete assessment of stress from the machine learning-based stress detection algorithm as the blue line denoted as $s(t)$. This moving average filter smooths the rapidly changing binary output of the stress detection algorithm in the time domain \cite{sarker2016finding}. The human condition metric is defined to be in the closed range $[0,1]$ for worst to perfect; therefore, the continuous stress plot $s(t) \in [1,0]$ is mapped to the estimated human condition as $1-s(t)$ to obtain a continuous human condition metric required in the MH-MR workload allocation framework. We chose this specific dataset in our study to show stress and subsequent human condition assessment because of the following properties: drastically changing human condition between $0-15$\,mins, slow change between $15-50$\,mins and sudden changes between $25-40$\,mins. We validate the effectiveness of our proposed framework on simplified cases of these rates of change of human condition in Section \ref{sec:s1}.

We deem noise reduction and disturbance rejection in measuring human operator condition as beyond the scope of our current work and included within the above presented human condition measurement metrics; a few specific works on signal processing and noise filtering of physiological measurements have been proposed in \cite{lahmiri2015physiological,li2007robust}. Therefore, we assume that human operator condition can be measured and quantified with enough certainty and noise rejection for application in our proposed MH-MR workload allocation framework.

To establish the generality of the modular human condition assessment function block in the proposed work allocation framework, we stress the following notes on possible human-operated robot scenarios. In cases where one human operates one robot or one human operates multiple robots, the human operator's health condition would be independently used in Module H for each of the operated robot's work allocation in Module A. However, if multiple humans control a single robot for an MH-MR application, the condition assessments of all the human operators of this particular robot would have to be considered for its workload allocation. In such a scenario where one robot is operated by more than one human operator, we assume that each human operator of the robot has exclusive on-board tasks: one operates navigation, one operates surveillance etc.; if one operator's condition deteriorates, one on-board task is affected. i.e. as a whole, this specific multi-human robot team's working ability is also affected. The definition of $\gamma$ in Eq. (\ref{eq:S_set}) ensures that the system allocates zero workload to a robot for any of its operators becoming incapacitated. The proposed MH-MR workload allocation system is therefore general to any number of operators controlling any number of robots in the system. With these generality notes, Eq. (\ref{eq:sigma}) assigns workload to individual robots reflecting its individual \textit{ability} to work considering the conditions of all its human operators.

\subsubsection{Module R: Robot Condition Evaluation Metric}
On-board quantitative measurements of robot \textit{health} may include battery level, communication signal strength, internal temperature and a variety of other factors \cite{reichard2004integrating}. Detecting sub-nominal characteristics and isolating problems through self-checking have also been considered in different autonomous robot platforms currently available. Qualitative evaluations may also be included for robot condition evaluation based on the robot's physical state. We refer to the Neglect Tolerance metric \cite{goodrich2003seven} for autonomous robots as a measure of how a robot's effectiveness declines in autonomous mode without any human supervision or control. It includes task complexity and robot capability among various other factors to provide an overall measure of a robot's condition of autonomy.

\subsubsection{Module P: Robot Performance Evaluation Metric}
Robot (either autonomous or human-operated) performance metrics such as percentage area coverage or distance travelled proposed in \cite{wong2002performance} may be used to asses robot performance depending on the MH-MR application. Performance of robots may be determined in terms of task completion time, path following cross-track error \cite{cruz2014navigation} etc. depending on the application of the proposed framework. Human Robot Interaction (HRI) metrics recommended by Steinfeld \textit{et al.} \cite{steinfeld2006common} and reviewed by Murphy \cite{murphy2013survey} in terms of navigation (e.g. localization, effective path determination around objects), perception (e.g. surveillance, target identification, sensor area coverage), manipulation, and management at the human, robot, and system level perspectives may be used to evaluate human-operated robot performance using an arbitrary evaluation function plugged in as Module P. We leave performance assessments for Module P at the discretion of their relevance to particular applications.

\section{Validation \& Results}\label{sec:val}
As validation of the effectiveness of the adaptive task allocation mechanism, we present our experimental findings of applying the proposed framework to a MH-MR patrolling application, where human operator and robot conditions affect their patrolling ability.

Four experiment scenarios were independently investigated. In the first scenario (S1), we simulate temporary and permanent deteriorated conditions for a human operator and an autonomous robot in sequence, while in the second scenario (S2) we simulate complete failure of a robot, and analyze the system's workload adaption in each scenario. In two further scenarios (S3 and S4) we present workload transitioning and scalability analysis with similar conditions as S1 and S2 respectively.

Before moving on to including real human operators in the experiments, it is vital that controllable evaluation scenarios are used to validate our proposed work. Therefore, in this paper we present the investigation results of our proposed method using simulated human operator conditions of different characteristics.

\subsection{S1 and S2: Adaptive Workload Allocation in Patrolling}
\subsubsection{Patrolling Application}\label{sec:patrol}
Machado \textit{et al.} broadly defined patrolling as ``the act of walking or traveling around an area, at regular intervals, in order to protect or supervise it" \cite{machado2002multi}. Therefore, we set up our representative patrolling application with a given number of robots traveling around allocated rectangular regions on a plane, where the sum of the area of all rectangular regions represents the total workload. The allocated workload from our proposed framework may be directly used in more complex region allocations for the patrolling scenario following capacity-constrained Voronoi tessellation works proposed in literature \cite{balzer2009capacity}. Applications in multi-robot coverage problems include \cite{karapetyan2017efficient}. However, for simplicity and ease of analysis we validate our system using rectangular patrolling regions, and we define patrolling for each robot as boundary following its allocated area within a specified time $\tau^*$ within its ability. 

A patrolling performance metric is defined for comparison study with and without the proposed workload allocation method. Patrolling performance of the complete MH-MR system is measured as the maximum time taken to patrol the entire area once by the MH-MR system expressed as,
\begin{equation}\label{eq:perform_metric}
T_L = max(t_{l_1}, t_{l_2}, .. , t_{l_m}) 
\end{equation}
where $t_{l_i}$ denotes the patrolling lap time of $R_i$ for $i \in I_R$ during one cycle of full area patrolling, given that the entire area is covered by all robots.
\begin{figure}[t]
    \centering
    \includegraphics[width=\linewidth]{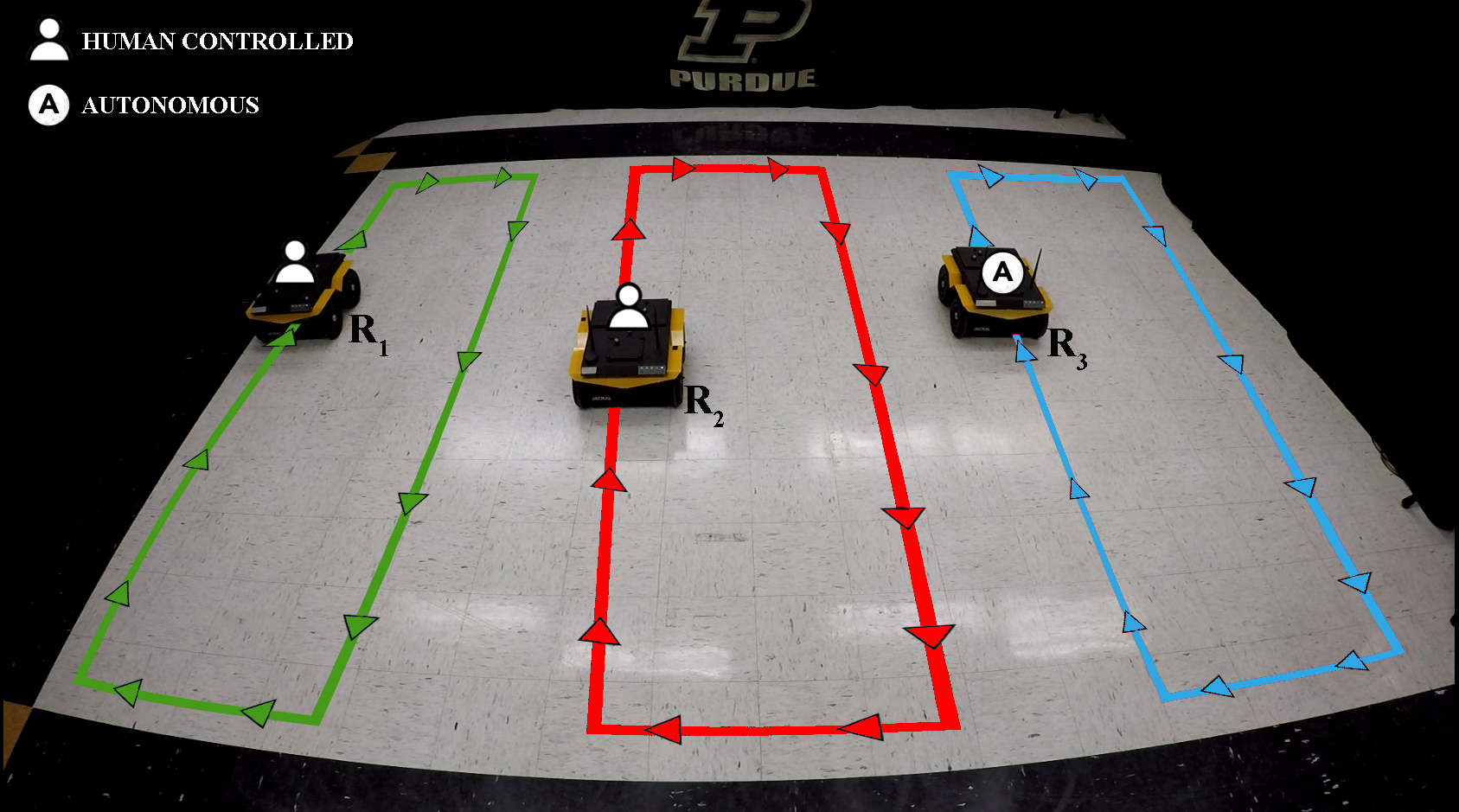}
    \caption{Experiment setup with $m=3$ Jackal mobile robots; $R_1$ and $R_2$ human-operated, and $R_3$ autonomous. Robots \textit{patrol} rectangular regions on the plane, defined as boundary following its allocated area. Patrolling velocities are modeled dependent on human operator and robot conditions.}
    \label{fig:expt_setup}
    \vspace{-5mm}
\end{figure}

\subsubsection{Experiment Setup}
We consider a MH-MR system of $h=2$ human operators (simulated) and $m=3$ mobile robots (Jackals from Clearpath Robotics) with position defined as $\mathbf{q}_i = [x_i, y_i]$, for $i \in I_R$ on a level plane as shown in Fig. \ref{fig:expt_setup}. Robot position data was recorded using a VICON system. True velocity estimation of the Jackals were made from the collected position data with time. We simulate robots $R_1$ and $R_2$ as being controlled by human operators while $R_3$ remains autonomous in patrolling. To simulate the human operators, human operator condition assessment inputs are provided for $R_1$ and $R_2$; all robots utilize the same low level line-of-sight path following controller for consistency. The effect of workload change in the system at time $t$ depends on the minimum distance from $q_i, \forall i \in I_R$ to the changing rectangular region boundaries using the allocation transition coefficient $K_e$ model in Eq. (\ref{eq:Ke}). 

At initial time, the patrolling area was distributed equally among all robots as rectangular regions with a specified safety distance between rectangular boundaries to prevent inter-robot collisions while patrolling, and the human operator and robot conditions were considered optimal. In course of the experiments, the rectangular region areas were re-allocated based on the proposed workload allocation framework. We acknowledge that increasing workload on an agent due to re-allocation, may reduce performance or in turn cause condition deterioration. Nevertheless, such effects on agents were ignored for validation purposes of the proposed framework. 

We model robot patrolling ability $v_{i}^{able}$ dependent on current human operator and robot condition,   
\begin{equation}\label{eq:vel_sim}
v_{i}^{able} = \kappa v_{max}
\end{equation}
for,
\begin{align}
\kappa = \begin{cases}
\text{min$(c_k^o, c_i^r)$} & i \in I_H, \forall k \in \Lambda_i\\
c_i^r & i \in I_A
\end{cases}
\end{align}
where $\Lambda_i$ is a vector of $\lambda \in I_O | \{R_i,O_{\lambda}\} \in E$, assuming $c_k^o$ and $c_i^r$ are bounded within $[0,1]$, and $v_{max}$ denotes the maximum allowed velocity of $R_i$. 

The required patrolling velocity of $R_i$ is modeled as,
\begin{equation}
    v_{i}^{req} = \frac{\text{Perimeter$(\mathcal{W}_i$})}{\tau^*} \quad \text{for $i={1,...,m}$}.
\end{equation}
where the patrolling time threshold is set as $\tau^*=65 \pm 10$\,s and $v_{i}^{req}=[0, v_{max}]$. Velocity of $R_i$ is therefore modeled as,
\begin{equation}
    v_{i} = \text{min}(v_{i}^{able},v_{i}^{req}).
\end{equation}
The initial value of $\tau^*$ is arbitrarily set large enough for experimental analysis purposes with $v_{max} = 0.8$\,m/s.

Validation setup parameters include the simulated human and robot condition update frequency time set as $\tau = 500$\,ms, workload transition scaling constant $K=0.5$.

Performance of each robot on the patrolling task is measured as cross-track error with an error margin of $\psi$. In reality, the performance measure would also include $v_i-v_i^{actual}$ corresponding to deteriorated performance of the robot. However, we intentionally do not consider velocity differences in our robot performance assessment in this validation setup, since we focus on independent analysis and assessment of the proposed workload allocation based on human operator and robot condition only. Performance measure of all robots is assumed to be unity at all times.

\subsubsection{S1: Adaptation to Deteriorated Conditions}\label{sec:s1}
\begin{figure}[t!]
    \begin{subfigure}{0.48\textwidth}
    \centering
        \includegraphics[width=\textwidth,height=1.3in]{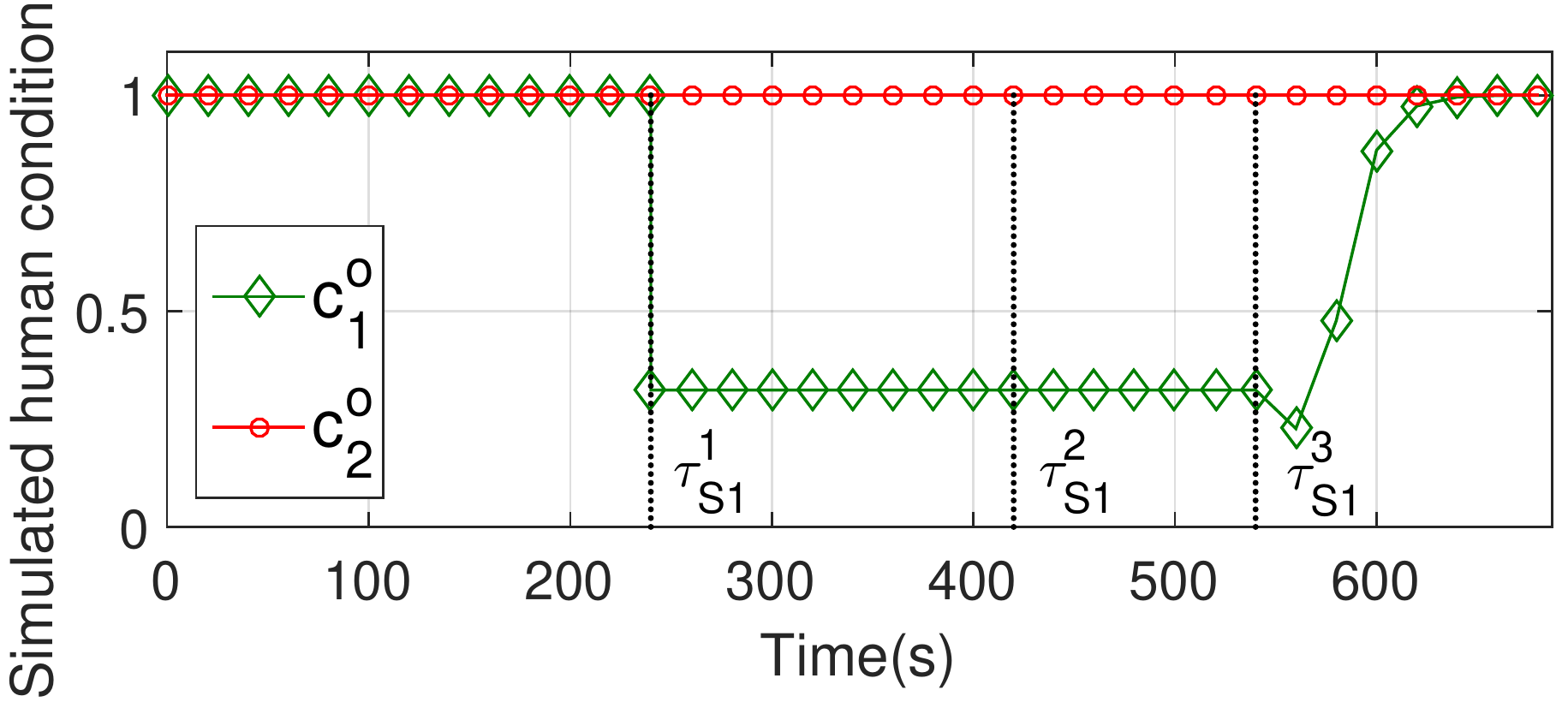}
        \caption{Simulated human conditions over time. $c_1^o$ deteriorates drastically at $\tau_{S1}^1$, and then subsequently recovers back to $1$ after a further small deterioration at $\tau_{S1}^3$; $c_2^o$ remains at $1$ at all times.}
        \vspace{1.5mm}
        \label{fig:c1_cond_h}
    \end{subfigure}
    \begin{subfigure}{0.48\textwidth}
    \centering
        \includegraphics[width=\textwidth,height=1.3in]{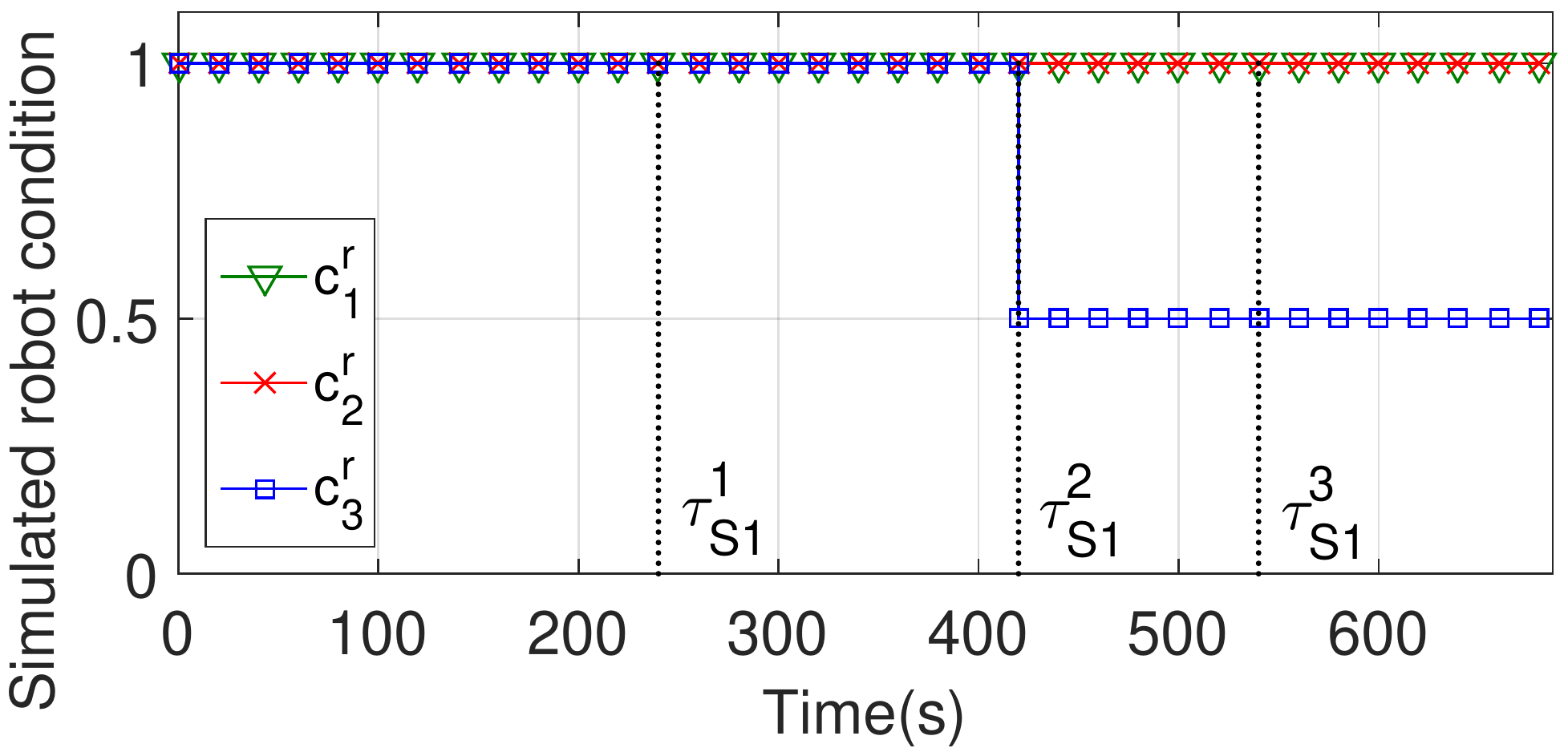}
        \caption{Simulated robot conditions over time. $c_3^r$ deteriorates permanently at $\tau_{S1}^2$; $c_1^r$, $c_2^r$ remains at $1$ at all times.}
        \vspace{1.5mm}
        \label{fig:c1_cond_r}
    \end{subfigure}
    \begin{subfigure}{0.48\textwidth}
    \centering
        \includegraphics[width=\textwidth,height=1.3in]{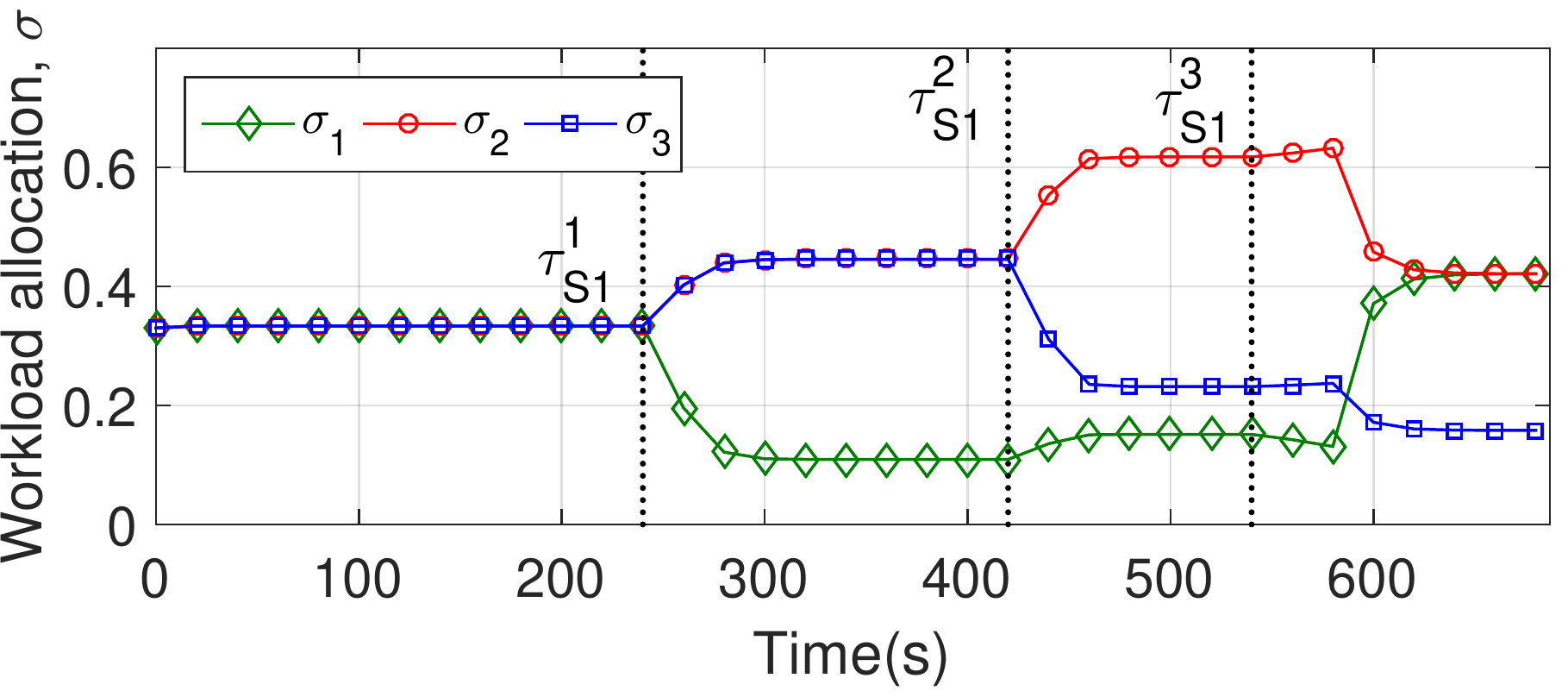}
        \caption{Workload allocation of patrolling robots change according to the simulated human and robot conditions: allocated workload of robots with deteriorated human and/or robot condition is reduced and compensated by robots with better human and/or robot condition. The workload transition function ensures that drastic changes are smoothened for a manageable effect on the host; yet remains sensitive enough to capture sudden changes in agent condition.}
        \vspace{1.5mm}
        \label{fig:c1_wl}
    \end{subfigure}
    \begin{subfigure}{0.48\textwidth}
    \centering
        \includegraphics[width=\textwidth,height=1.5in]{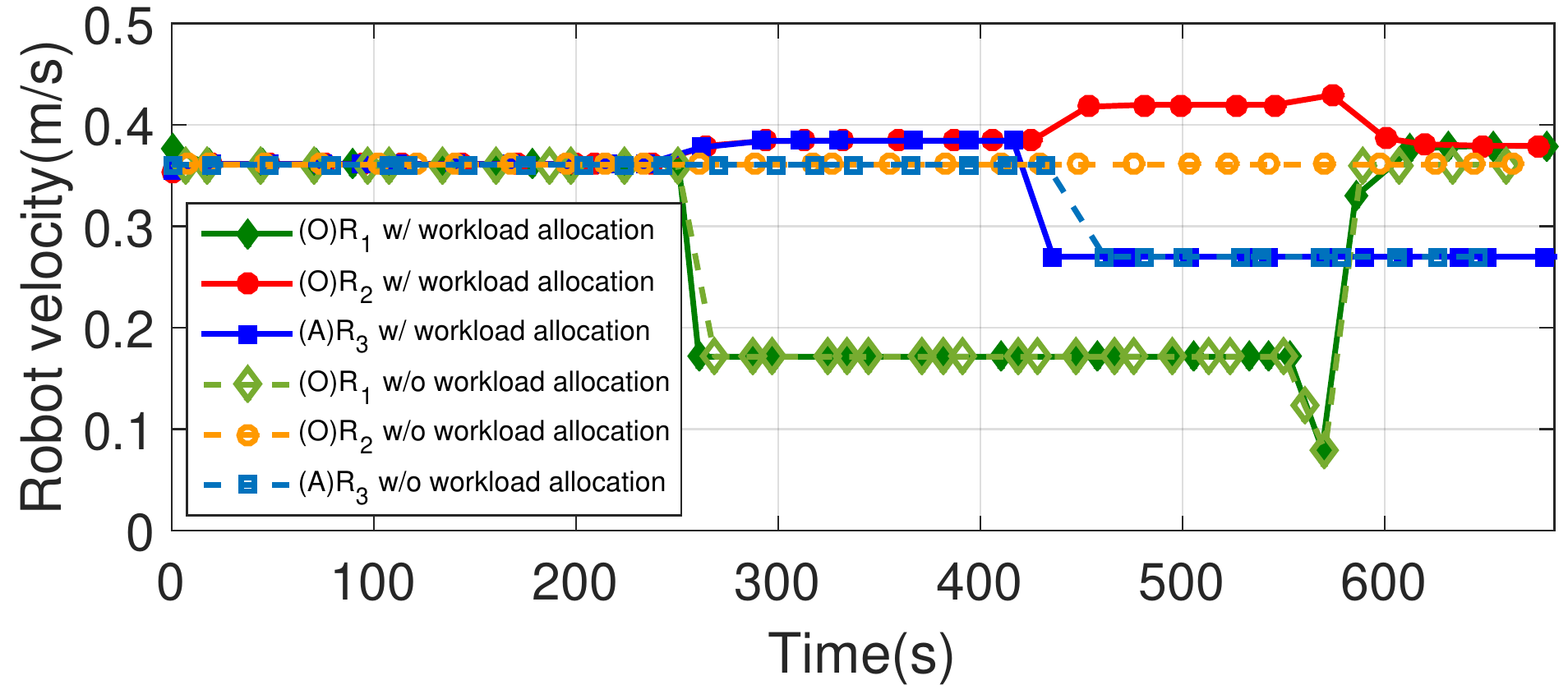}
        \caption{Translation velocity profiles of patrolling robots with and without workload allocation ignoring region corner rotations. Translation velocity of robots with deteriorated human and/or robot conditions is observed to have lowered velocity ($R_1$, $R_2$ after $\tau_{S1}^1$ and $\tau_{S1}^2$ respectively); robots with increased workload after re-allocation having better conditions observed to increase their translation velocity ($R_2$, $R_3$ after $\tau_{S1}^1$ and $R_3$ after $\tau_{S1}^2$).}
        \vspace{1.4mm}
        \label{fig:c1_vel}
    \end{subfigure}
    \caption{S1: Adaptive workload allocation for temporary and permanent human and robot condition deterioration.}
    \label{fig:c1_master}
    \vspace{-6mm}
\end{figure}
\begin{figure*}[t!]
    \centering
    \begin{subfigure}{0.45\textwidth}
    \centering
        \includegraphics[width=\linewidth]{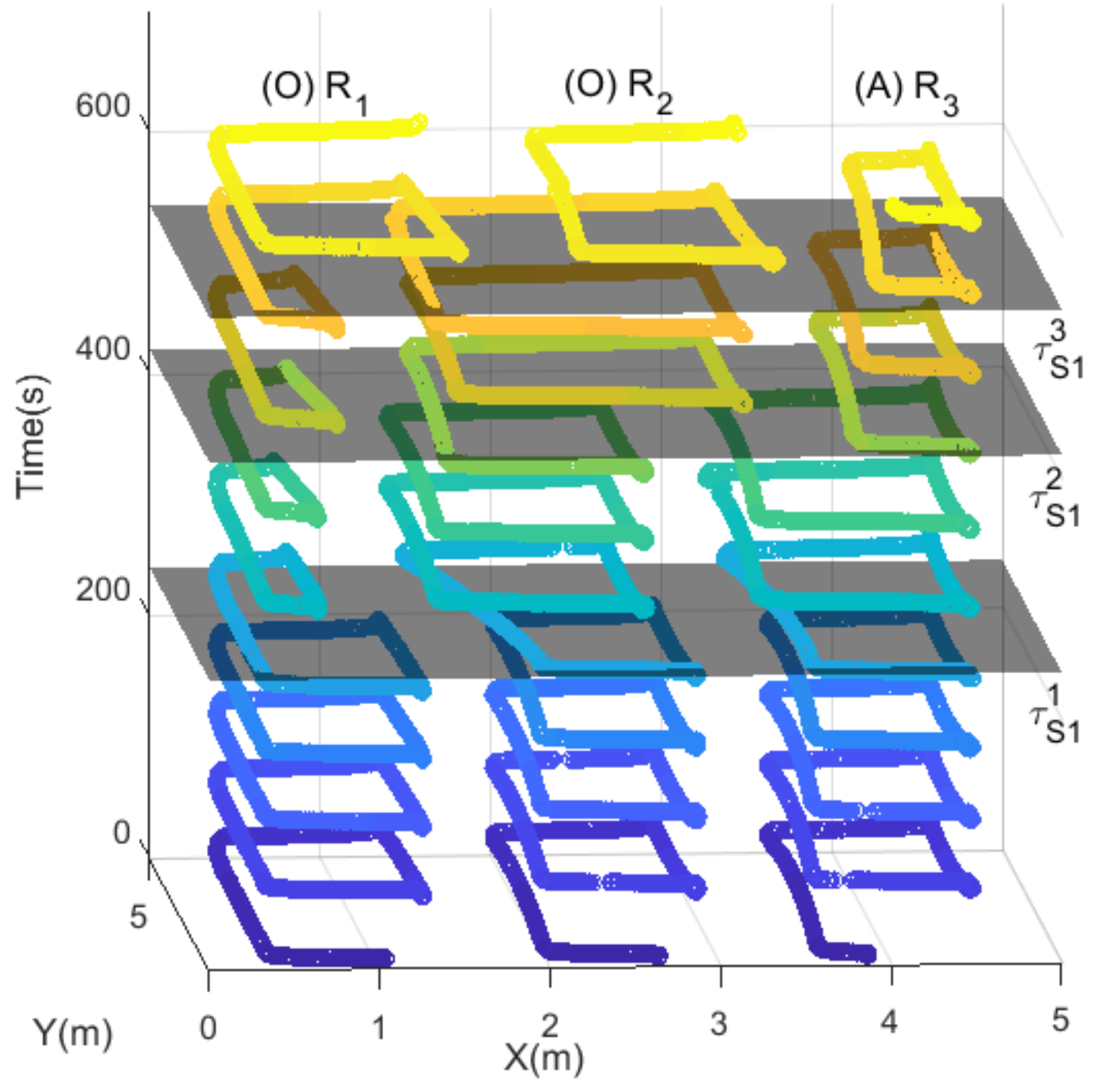}
        \vspace{-0.5cm}
        \caption{Patrolling trajectory of robots with workload allocation}
        \label{fig:c1_withtraj}
    \end{subfigure}\qquad
    \begin{subfigure}{0.45\textwidth}
    \centering
        \includegraphics[width=\linewidth]{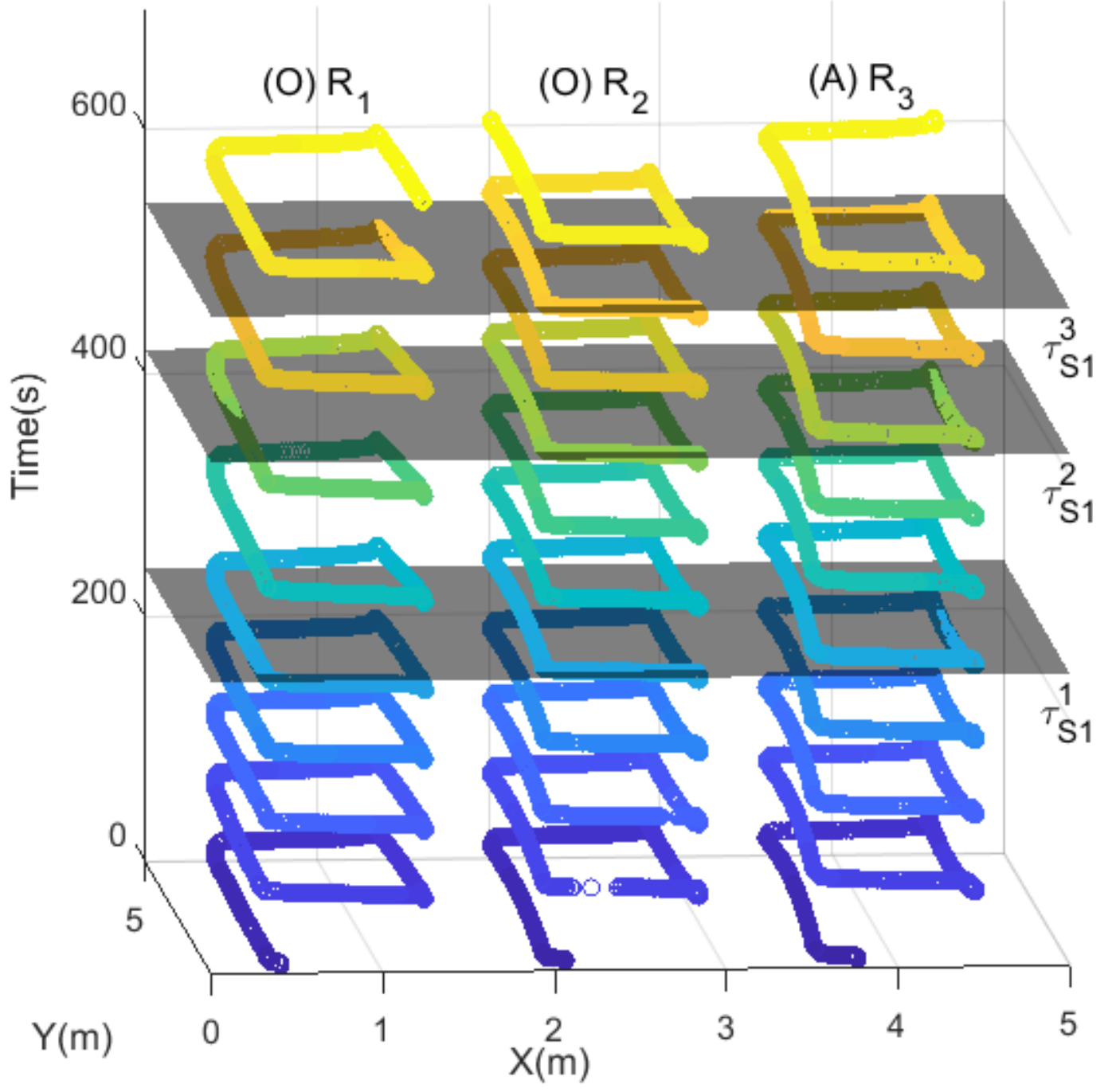}
        \caption{Patrolling trajectory of robots without workload allocation} 
        \label{fig:c1_wotraj}
    \end{subfigure}
    \caption{S1: Patrolling trajectory following comparison with and without workload allocation for temporary and permanent human and robot condition deterioration.}
    \label{fig:c1_master2}
    \vspace{-6mm}
\end{figure*}
\begin{figure}
    \centering
    \includegraphics[width=\linewidth,height=2.0in]{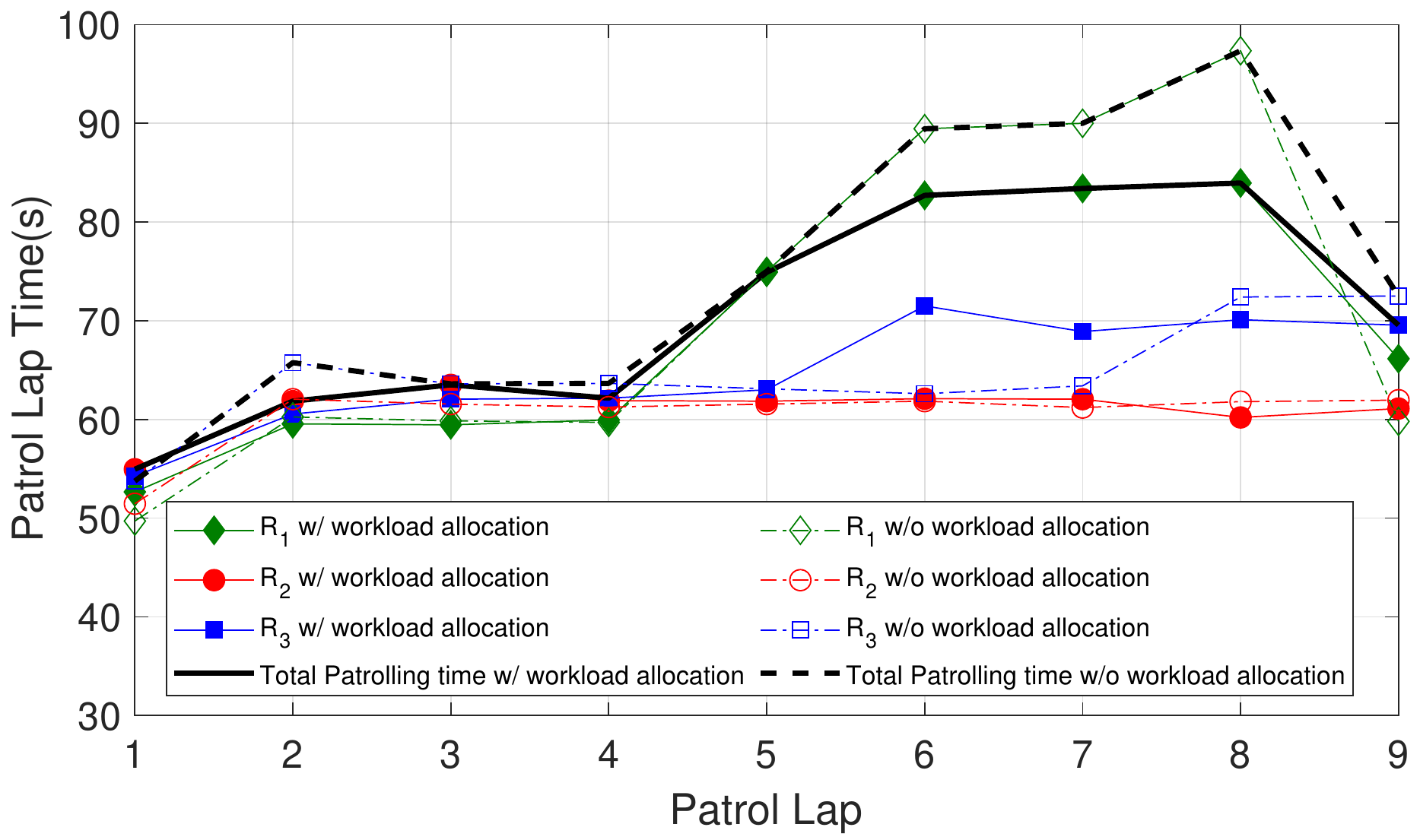}
    \caption{S1: Total and individual patrolling time required comparison with and without workload allocation.}
    \label{fig:c1_laptime}
    \vspace{-6mm}
\end{figure}

We model the conditions for the human operator $O_1$ of $R_1$ denoted as $c_1^o$ to deteriorate drastically at time $\tau_{S1}^1$ and then subsequently recover slowly back to $1$ after a sudden further small deterioration at time $\tau_{S1}^3$ as shown in Fig. \ref{fig:c1_cond_h}; this is based on simplified, observed and analyzed condition patterns of quantified human stress from Fig. \ref{fig:stress_detection_ML} (drastically changing human condition between $0-15$\,mins, slow change between $15-50$\,mins and sudden changes between $25-30$\,mins). Here we stress the design of the simulated events having drastic and different rates of changes on the two separate time instances (abrupt and slow), to show their effects on the workload allocation. The minor further deterioration before recovery after $\tau_{S1}^2$ is simulated to investigate the sensitivity of the proposed workload allocation framework to sudden small changes in operator condition. Deteriorated condition of the autonomous robot $R_3$ denoted as $c_3^r$ is simulated as shown in \ref{fig:c1_cond_r}. $c_3^r$ is simulated to deteriorate permanently at $\tau_{S1}^2$. The experiment S1 is repeated with and without the proposed adaptive workload allocation framework to compare their effects on the patrolling application using the defined patrolling performance metric. The results are presented in Fig. \ref{fig:c1_master}, \ref{fig:c1_master2} and \ref{fig:c1_laptime}.

With initially set equal workload, all robots patrol their equally allocated rectangular boundaries until time event $\tau_{S1}^1$. At $\tau_{S1}^1$ where $c_1^o$ shows drastically falling conditions, the workload is re-allocated to reduce load on the human $O_1$ operated robot $R_1$ and increased equally among $R_2$ and $R_3$ having better conditions as seen in Fig. \ref{fig:c1_wl}; the re-allocation is reflected as a smaller patrolling region for $R_1$ and equal larger regions for $R_2$ and $R_3$ in Fig. \ref{fig:c1_withtraj}. $R_2$ and $R_3$ were both positioned roughly equally close to the changing boundary of their rectangular regions during the first event at $\tau_{S1}^1$, and thus both robots transition to their allocated workload at the same time of around $300$\,s shown in Fig. \ref{fig:c1_wl}. The corresponding changes in the velocity profiles for each robot with workload allocation is shown in Fig. \ref{fig:c1_vel}. The velocity of $R_1$ is seen drastically reduced with $v_1^{able}<v_1^{req}$; and with increased allocations of patrolling regions, the other two robots at this point still in good condition are observed to slightly increase their velocities ($v_2^{req}<v_2^{able}$ and $v_3^{req}<v_3^{able}$). The patrolling time for $R_1$ is observed as increasing above the $\tau^*$ tolerance at lap $5$ and eventually levelling at lap $6$ due to the slow workload transition process as shown in Fig. \ref{fig:c1_laptime}; and remained high over laps $6$ and $7$ due to more frequent slower turning at corners. Patrolling times for $R_2$ and $R_3$ with optimal conditions remained steady within defined $\tau^*$ tolerance after time event $\tau_{S1}^1$. 

Similar observations are made after time event $\tau_{S1}^2$, where robot $R_3$ suffers a sudden condition deterioration. The workload of $R_3$ is reduced and re-distributed among the other two as seen in Fig. \ref{fig:c1_wl} and \ref{fig:c1_withtraj}. Velocity of $R_3$ decreases permanently due to its deteriorated condition. $R_2$ is left with patrolling a larger region in comparison to the others; its velocity increases to maintain the patrolling time requirement. However, the velocity of $R_1$ remains the same due to its previously deteriorated condition. Thus, the patrolling time for $R_1$ remains considerably higher than $R_2$ and $R_3$ for laps $7$ and $8$ after time event $\tau_{S1}^2$ as shown in Fig. \ref{fig:c1_laptime}. 

Compared to the drastic change of $c_1^o$ at $\tau_{S1}^1$ and $c_3^r$ at $\tau_{S1}^2$, $c_1^o$ gradually returns to $1$ after a sudden drop at $\tau_{S1}^3$. The workload allocation is seen to change relatively slowly as well for this time event over time period $540$\,s to $630$\,s as observed in Fig. \ref{fig:c1_wl}. Right before the recovery, the workload allocation plot shows a slight dip in allocated workload for $R_1$ and small increases for $R_2$ and $R_3$ validating the effective sensitivity of the proposed workload allocation method. Upon condition improvements of $O_1$ at $\tau_{S1}^3$, the workload is redistributed again to equal portions among $R_1$ and $R_2$ with corresponding rectangular block patrolling trajectories shown in Fig. \ref{fig:c1_withtraj}. The velocity of $R_1$ and $R_2$ equalize to a larger value than $R_3$ to compensate for the re-allocated patrolling regions with lower workload for $R_3$. Patrolling lap times for all robots return within the defined $\tau^*$ tolerance after event $\tau_{S1}^3$ at patrol lap $9$ with the highest time taken by $R_3$.
\begin{figure}[t!]
    \centering
    \begin{subfigure}{0.48\textwidth}
    \centering
        \includegraphics[width=\textwidth,height=1.3in]{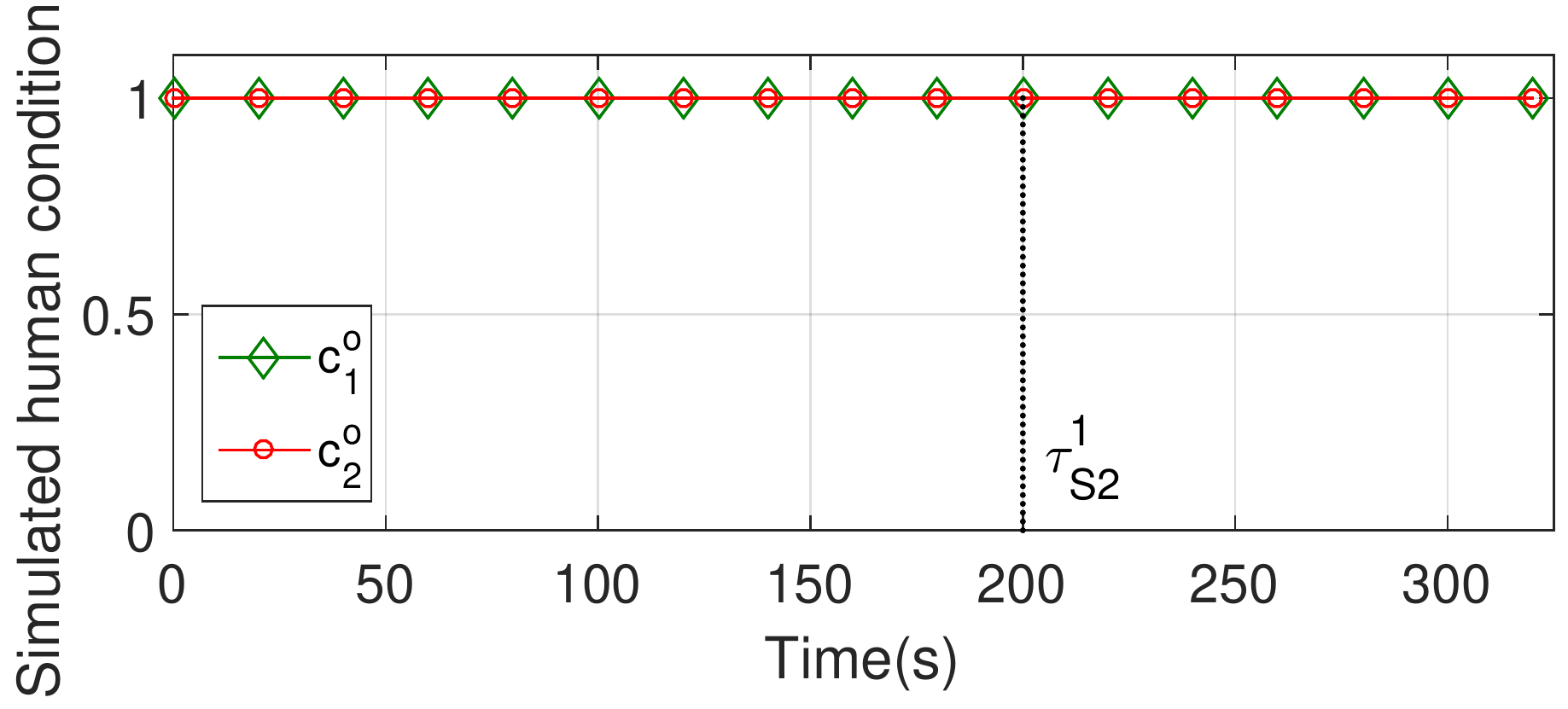}
        \caption{Simulated human conditions over time. $c_1^o$ and $c_2^o$ remains at $1$ at all times.}
       \vspace{0.9mm}
        \label{fig:c2_cond_h}
    \end{subfigure}
    \begin{subfigure}{0.48\textwidth}
    \centering
        \includegraphics[width=\textwidth,height=1.3in]{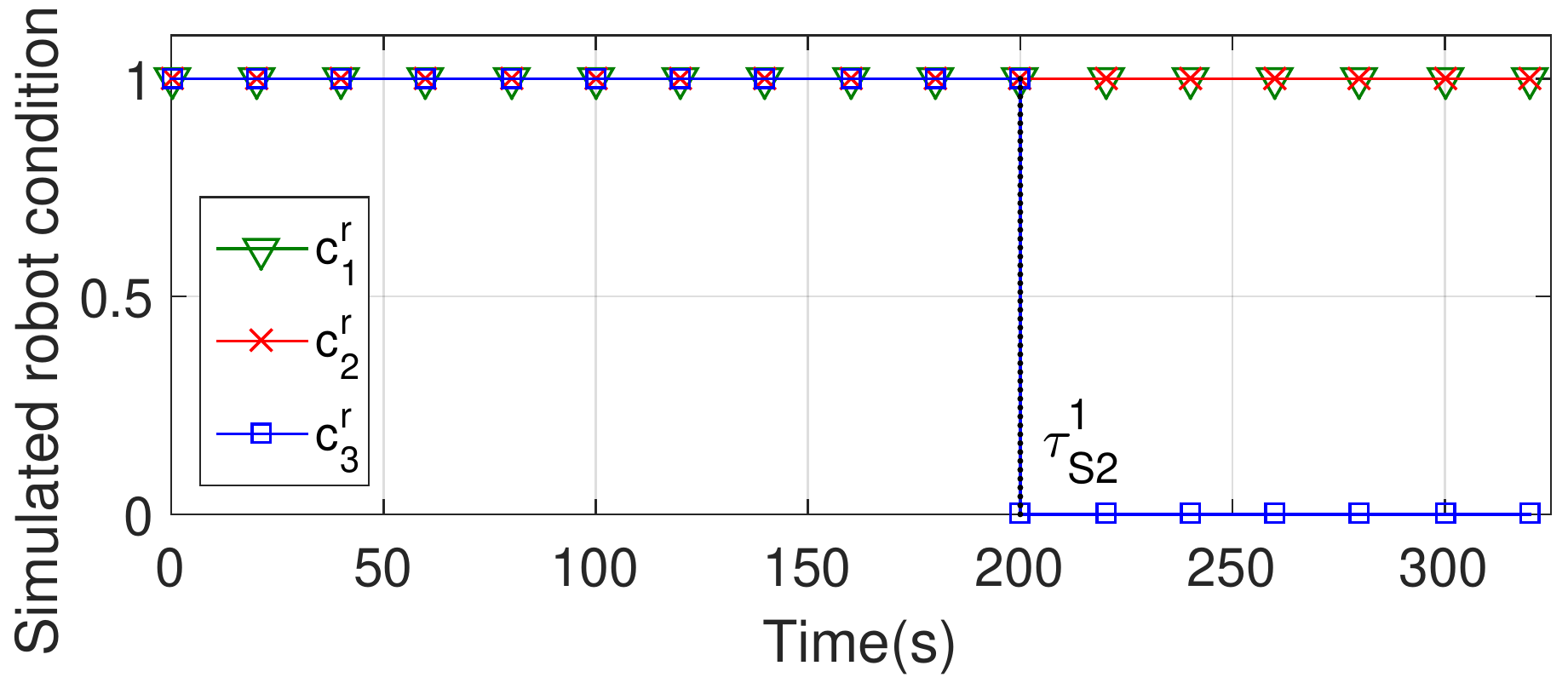}
        \caption{Simulated robot conditions over time. $c_1^r$ and $c_2^r$ remains at $1$ at all times; $c_3^r$ deteriorates permanently at $\tau_{S2}^1$.}
       \vspace{0.9mm}        
        \label{fig:c2_cond_r}
    \end{subfigure}
    \begin{subfigure}{0.48\textwidth}
    \centering
        \includegraphics[width=\textwidth,height=1.3in]{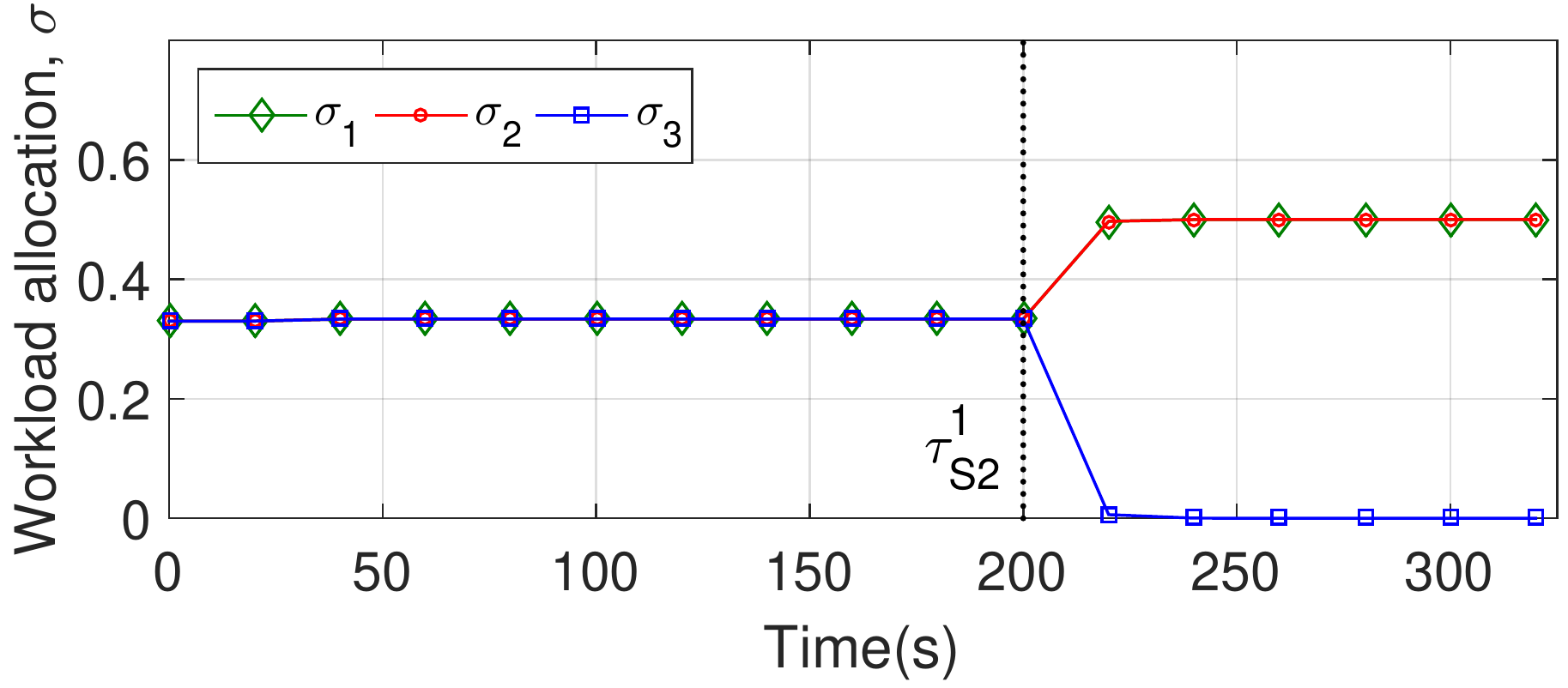}
        \caption{Workload allocation of patrolling robots change according to the simulated human and robot conditions: allocated workload of incapacitated robots is zero and compensated by robots with better human and/or robot condition. The workload transition function module ensures that drastic changes are smoothened for a manageable effect on the host.} 
       \vspace{0.9mm}        
        \label{fig:c2_wl}
    \end{subfigure}
    \begin{subfigure}{0.48\textwidth}
    \centering
      \includegraphics[width=\textwidth,height=1.3in]{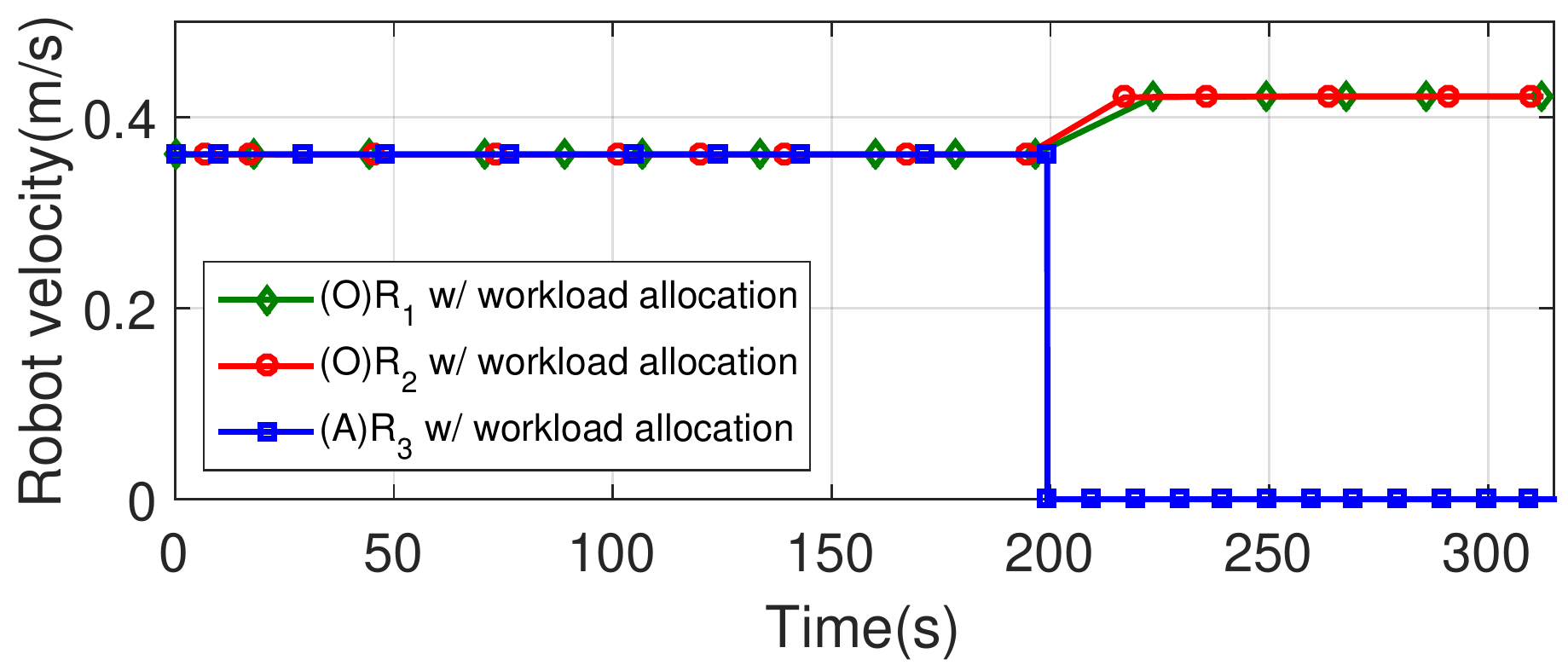}
      \caption{Translation velocity profiles of patrolling robots with and without workload allocation ignoring region corner rotations. Translation velocity of incapacitated robots (deteriorated \textit{ability}) observed to be zero ($R_3$ after $\tau_{S2}^1$), while robots with increased workload after re-allocation having better conditions observed to increase their translation velocity to compensate ($R_1$, $R_2$ after $\tau_{S2}^1$).}
      \label{fig:c2_vel}
    \end{subfigure}
    \caption{S2: Adaptive workload allocation for complete robot failure condition.}
    \label{fig:c2_master}
    \vspace{-6mm}
\end{figure}
\vspace{-1mm} 

To validate the effectiveness of the proposed method, the experiment scenario is repeated without using the adaptive workload allocation framework. The robot patrolling trajectories followed the initial equal rectangular region allocation throughout the experiment as presented in Fig. \ref{fig:c1_wotraj}. With equal rectangular region allocation over the entire experiment duration, the robot velocities only reflected the temporary and permanent deteriorating conditions of $O_1$ ($v_1^{able}<v_1^{req}$) and $R_3$ ($v_3^{able}<v_3^{req}$) showing slower patrolling speed as observed in Fig. \ref{fig:c1_vel} and higher patrolling times after $\tau_{S1}^1$ and $\tau_{S1}^3$ respectively. Referring to the previously defined patrolling performance metric, the total area patrolling time was recorded to be $7$\,s higher on lap $5$ (right after $\tau_{S1}^1$) without workload allocation. With the initial dip in $c_1^o$ after $\tau_{S1}^3$, the area patrolling time was initially recorded to be $14$\,s higher on lap $8$ without workload allocation in comparison, that reduced within the set $\tau^*$ tolerance on lap $9$, when the simulated $c_1^o$ gradually returned to $1$. 

The events $\tau_{S1}^1$, $\tau_{S1}^2$ and $\tau_{S1}^3$ triggered changes in workload on immediate neighbors of $R_2$, allowing it to quickly adjust its velocity to meet the required patrolling lap time. The workload change after event $\tau_{S1}^1$, was slower for $R_3$ in comparison as it adjusted to the change following transitioning of $R_2$, resulting in increased patrolling lap times in laps $5$ and $6$. The event $\tau_{S1}^2$ triggered in between patrolling laps $6$ and $7$ of $R_3$ permanently kept its patrolling velocity at $70$\,s with workload allocation. In comparison, its patrolling lap time without workload allocation is observed to increase on lap $7$ and permanently stay $3$\,s higher for the rest of the experiment. Although insignificant compared to $R_1$, the total area patrol time remained $3$\,s less due to $R_3$ with the proposed workload allocation after event $\tau_{S1}^3$ on lap $9$.

\subsubsection{S2: Adaptation to Robot Failure}
Experimental cases of complete robot failures have also been investigated, where $R_3$ is completely incapacitated by setting $c_3^r = 0$ at event $\tau_{4}$ in a separate experiment. Fig. \ref{fig:c2_cond_h} and \ref{fig:c2_cond_r} shows the simulated human and robot conditions for S2. 

\begin{figure}[t]
    \centering
    \includegraphics[width=0.85\linewidth, height = 3in]{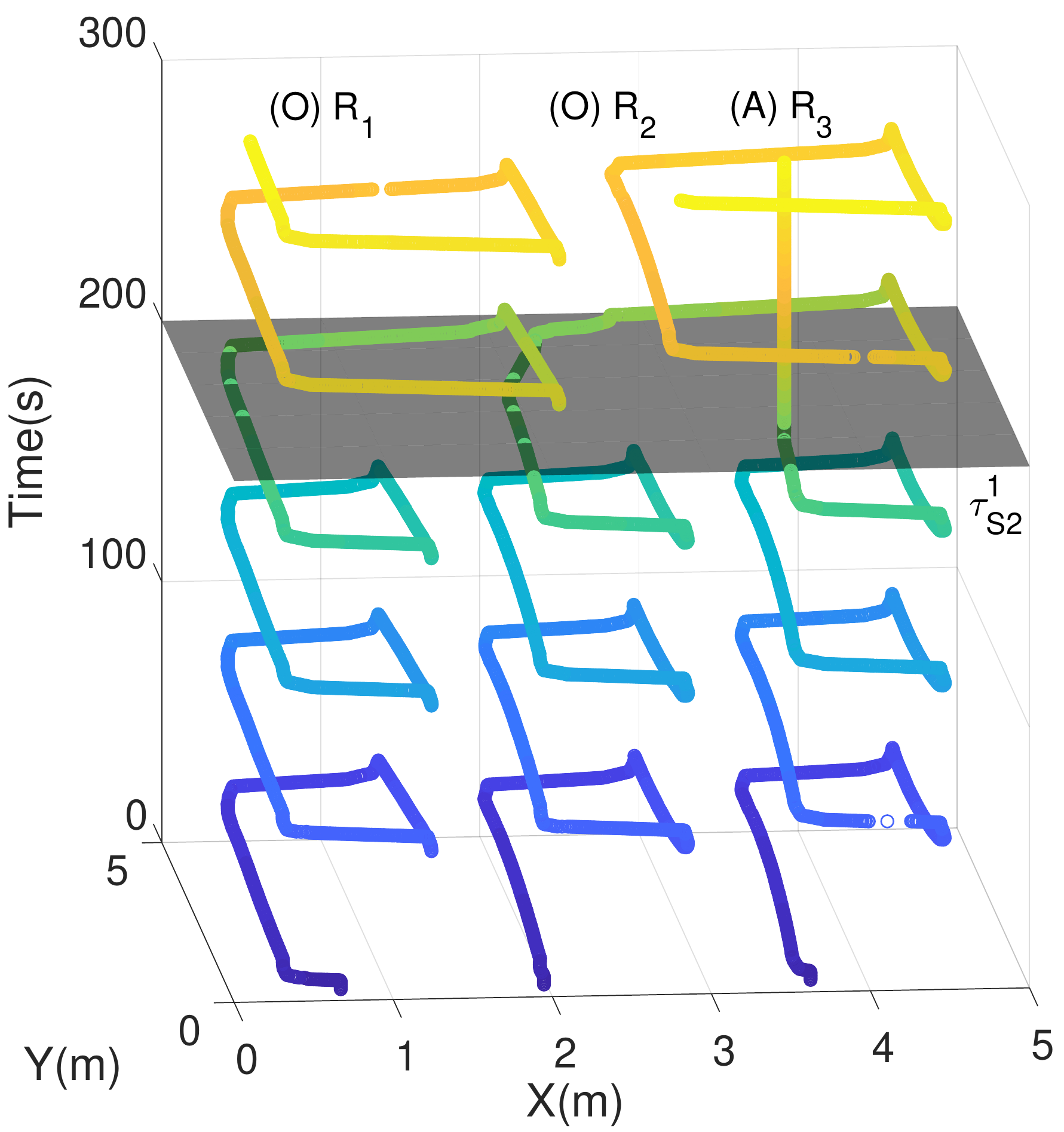}
    \caption{S2: Patrolling trajectory of robots with workload allocation for complete robot failure condition.}
    \vspace{-6mm}
    \label{fig:c2_traj}
\end{figure}
We refer to Fig. \ref{fig:c2_wl} to present the resulting allocated workloads after event $\tau_4$. At event $\tau_4$, the initial area of $R_3$ is equally allocated amongst $R_1$ and $R_2$ for continuous full patrolling area coverage; i.e. at any time instant, the total allocated workload was always unity with the proposed adaptive workload allocation framework. This verifies that the workload was always re-allocated to ensure total area coverage by the MH-MR patrolling system. The resulting robot trajectory plots are shown in Fig. \ref{fig:c2_traj}. The modeled velocity plots shown in Fig. \ref{fig:c2_vel} confirm the increased patrolling velocities for $R_1$ and $R_2$ to compensate for their allocated larger equal areas. For S2, we omit comparison of patrolling performance with and without using adaptive workload allocation, since total area patrolling could only be achieved with the proposed adaptive workload allocation framework. 
\begin{figure}[t]
    \centering
    \includegraphics[width=\linewidth,height=0.23in]{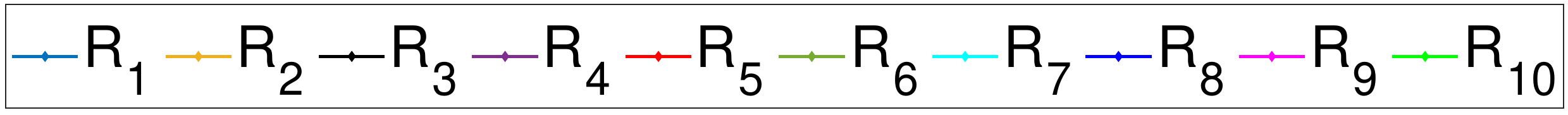}
    \vspace{4mm}
    \centering
    \sbox{\bigimage}{
    \begin{subfigure}[t]{0.235\textwidth}
        \centering
        \includegraphics[width=\textwidth, height=2.45in]{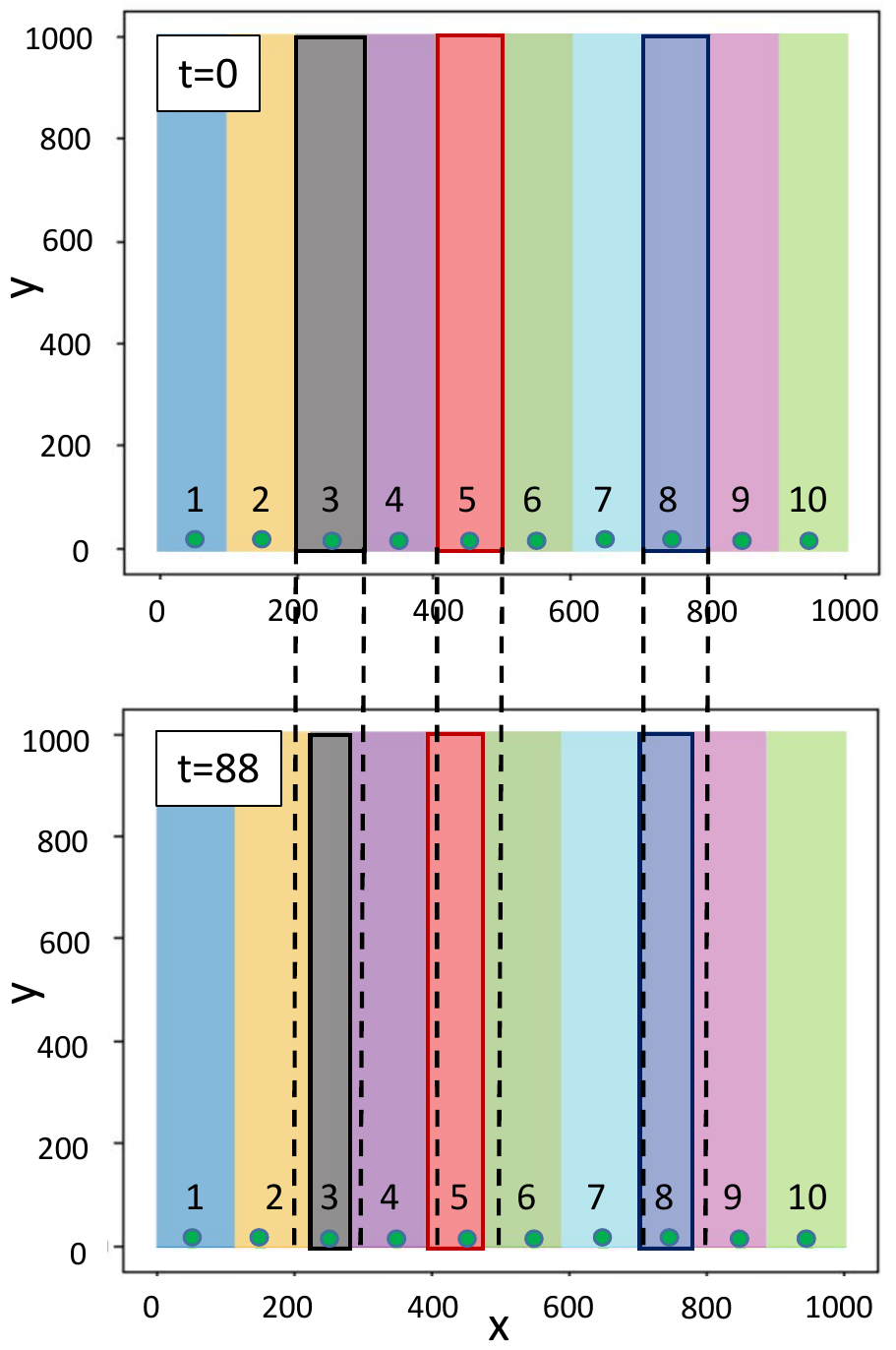}
        \caption{Initial and final workload allocated regions for $m=10$ stationary robot MH-MR system.} 
        \label{fig:c31_area}
    \end{subfigure}
    }
    \usebox{\bigimage}\hfill
    \begin{minipage}[b][\ht\bigimage][s]{.23\textwidth}
        \begin{subfigure}[t]{\textwidth}
        \centering
            \includegraphics[width=1.6in, height=1.1in]{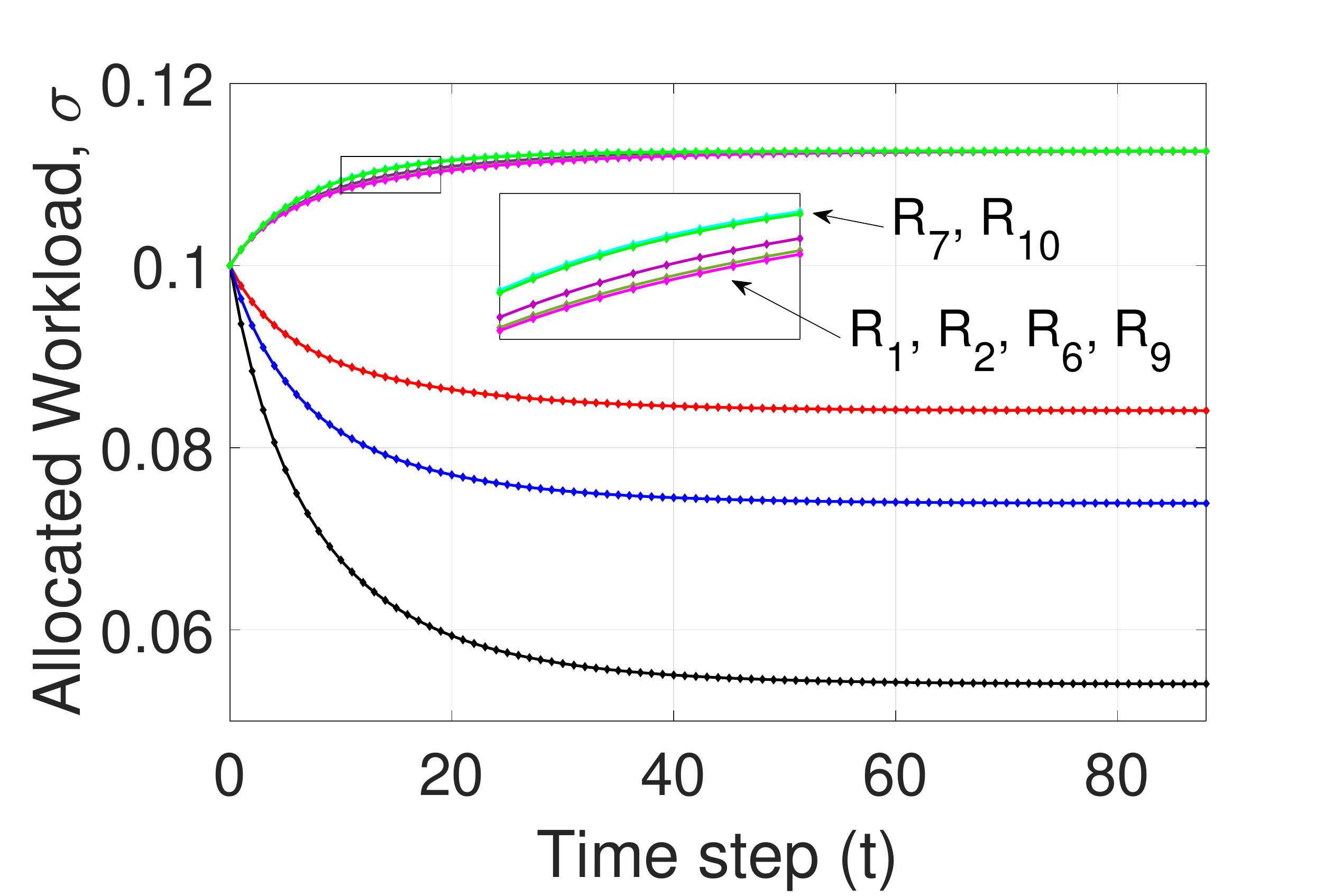}
            \caption{Allocated workload convergence for $K = 5$.}
            \label{fig:c31_allocConv}
        \end{subfigure}
        \vspace{-1.8mm}
        \begin{subfigure}[t]{\textwidth}
        \centering
          \includegraphics[width=\textwidth, height=1.0in]{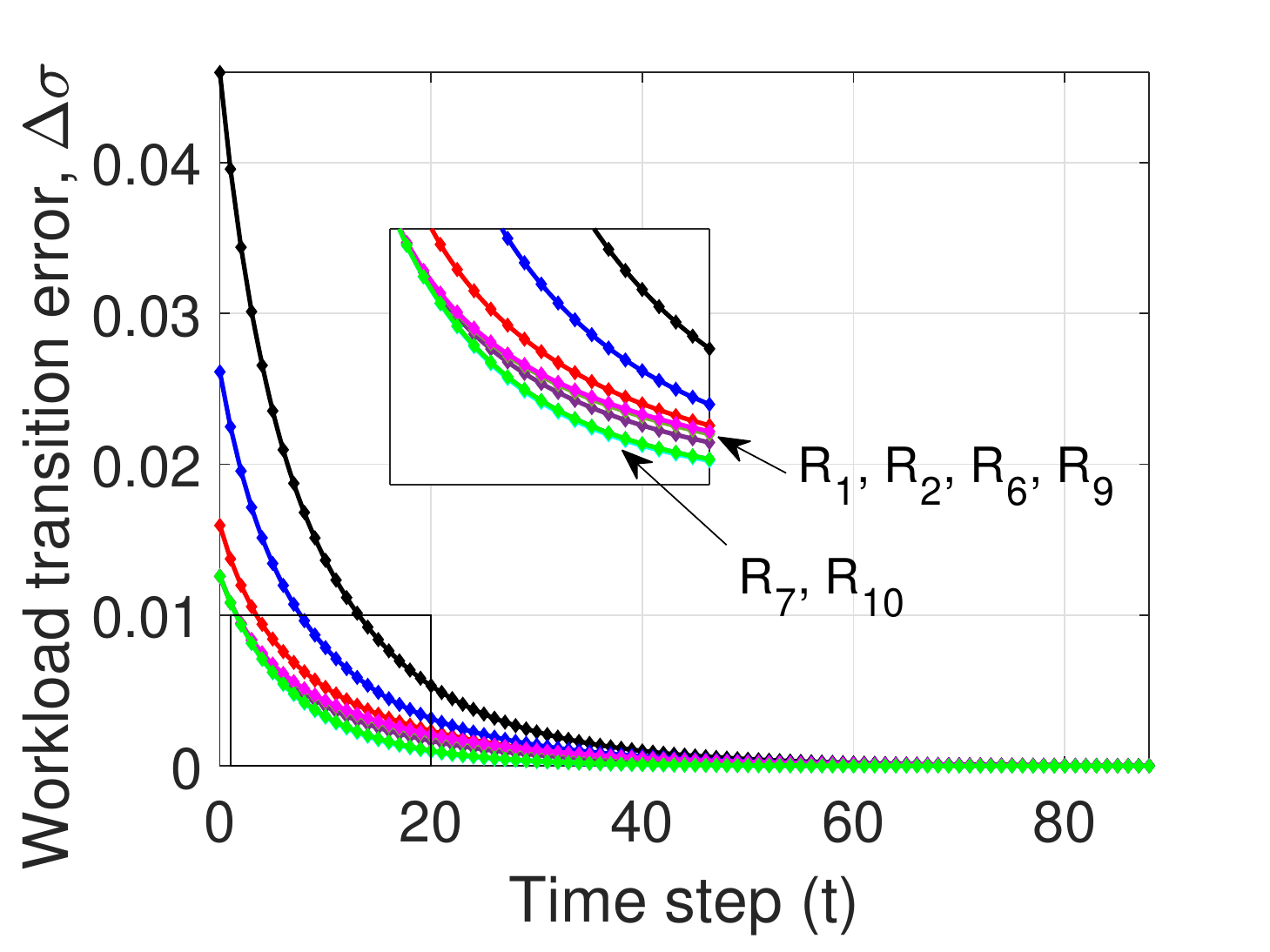}
          \caption{Allocated workload convergence error over time for $K=5$.}
          \label{fig:c31_allocError}
        \end{subfigure}
    \end{minipage}
    \vspace{-4mm}
    \caption{S3: Transition analysis for $m=10$ stationary robots equidistant from their region boundaries along the horizontal axis, with equal initial workload allocation. Human and robot conditions are simulated as $c^o_3=0.8$, $c^r_3=0.6$, $c^o_5=0.8$, $c^o_8=0.75$ with the rest as $1$ from $t=0$, and the system adaptively converges to the new workload depending on $q_f$. Green dots represent robots. Zoomed sections of plots shown in insets.}
    \label{fig:scalable_1}
    \vspace{-4mm}
\end{figure}

\subsection{S3 and S4: Workload Allocation Transition Analysis}\label{sec:scalable}
\begin{figure}[t]
    \centering
    \includegraphics[width=\linewidth,height=0.23in]{legend.pdf}
    \vspace{4mm}
    \centering
    \sbox{\bigimage}{
    \begin{subfigure}[t]{0.235\textwidth}
        \centering
        \includegraphics[width=\textwidth, height=2.45in]{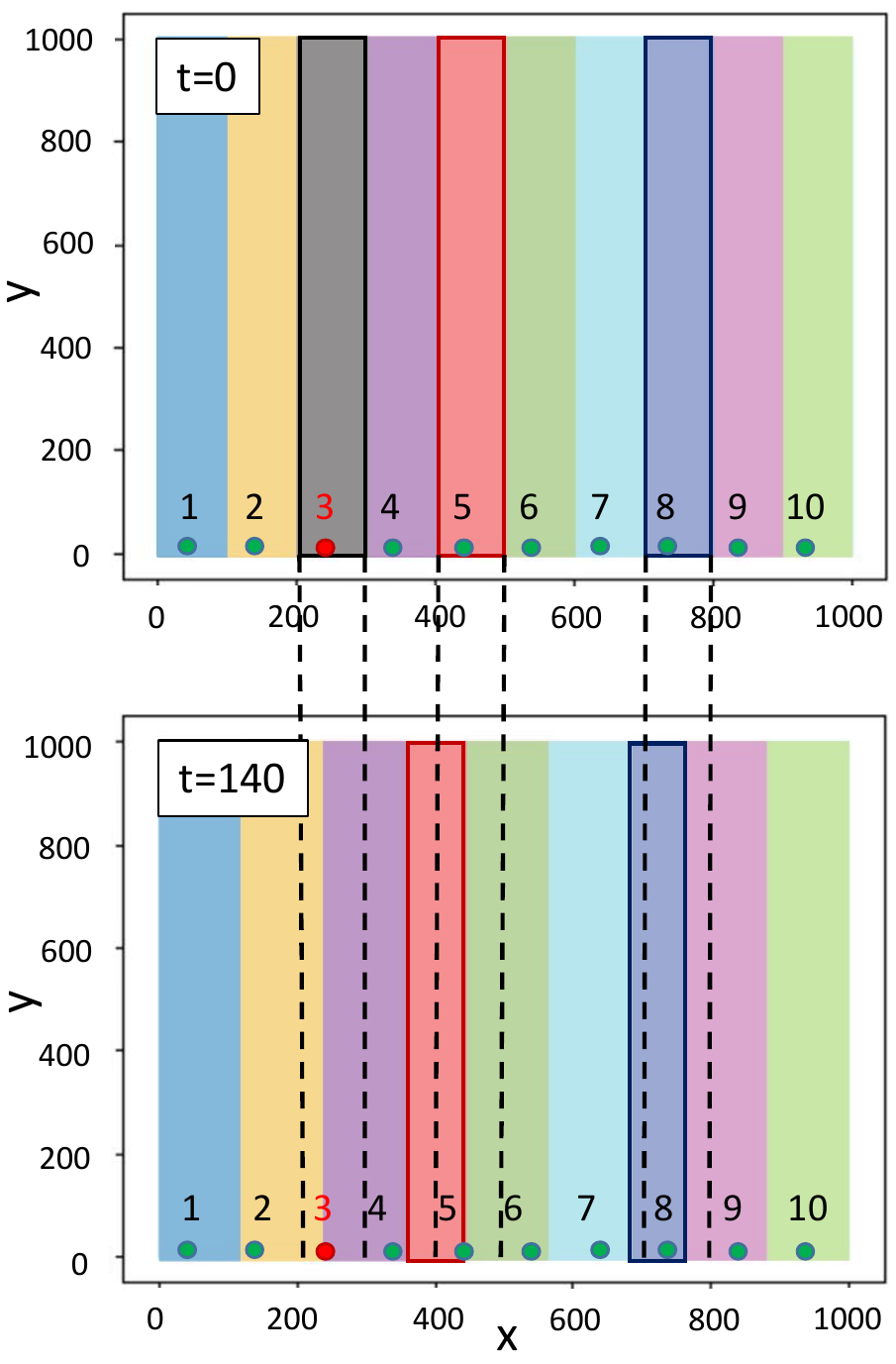}
        \caption{Initial and final workload allocated regions for $m=10$ stationary robot MH-MR system.} 
        \label{fig:c32_area}
    \end{subfigure}
    }
    \usebox{\bigimage}\hfill
    \begin{minipage}[b][\ht\bigimage][s]{.23\textwidth}
        \begin{subfigure}[t]{\textwidth}
        \centering
            \includegraphics[width=1.6in, height=1.1in]{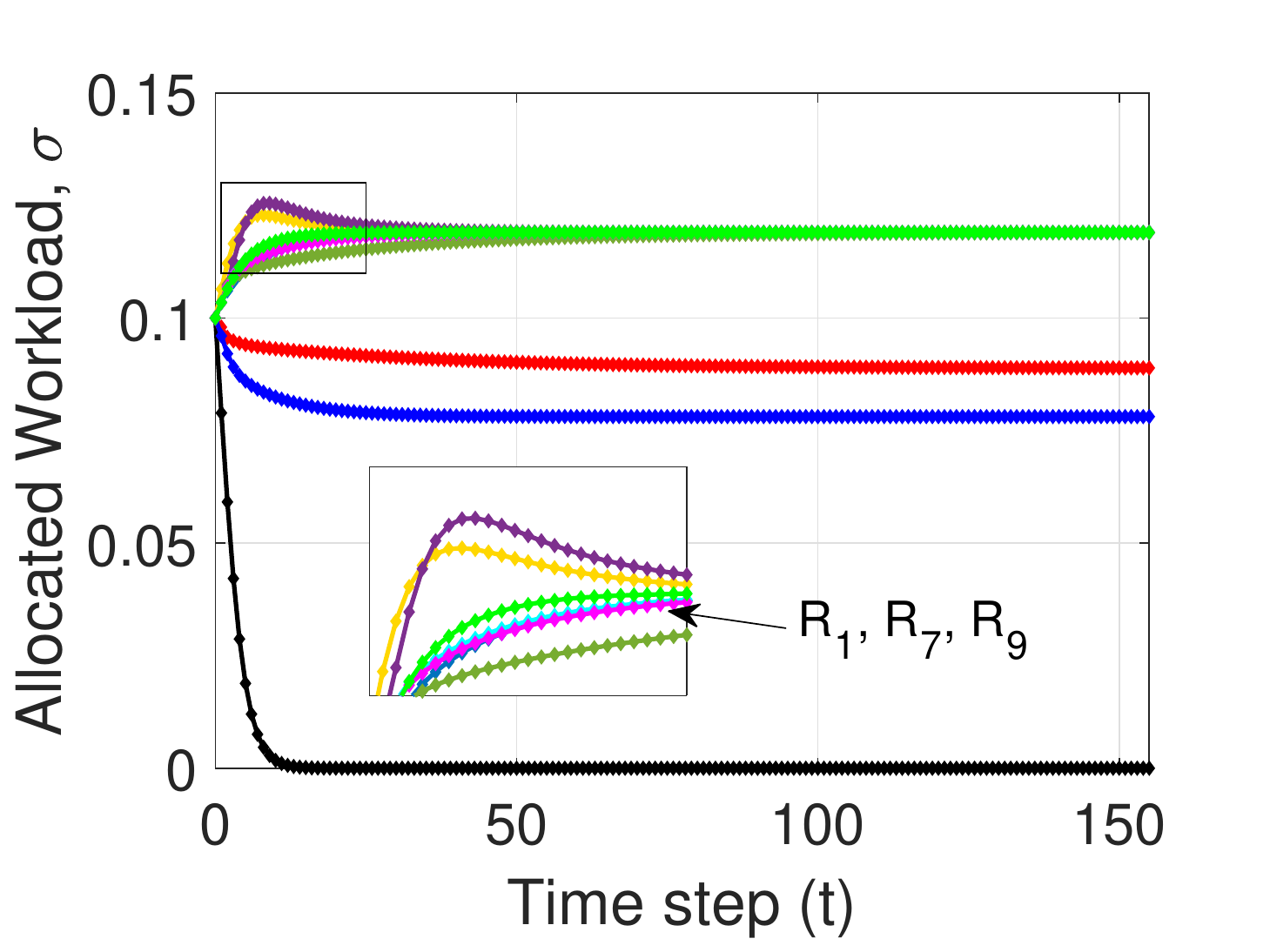}
            \caption{Allocated workload convergence for $K=5$.}
            \label{fig:c32_allocConv}
        \end{subfigure}
        \vspace{-1.8mm}
        \begin{subfigure}[t]{\textwidth}
        \centering
        \includegraphics[width=\textwidth, height=1.0in]{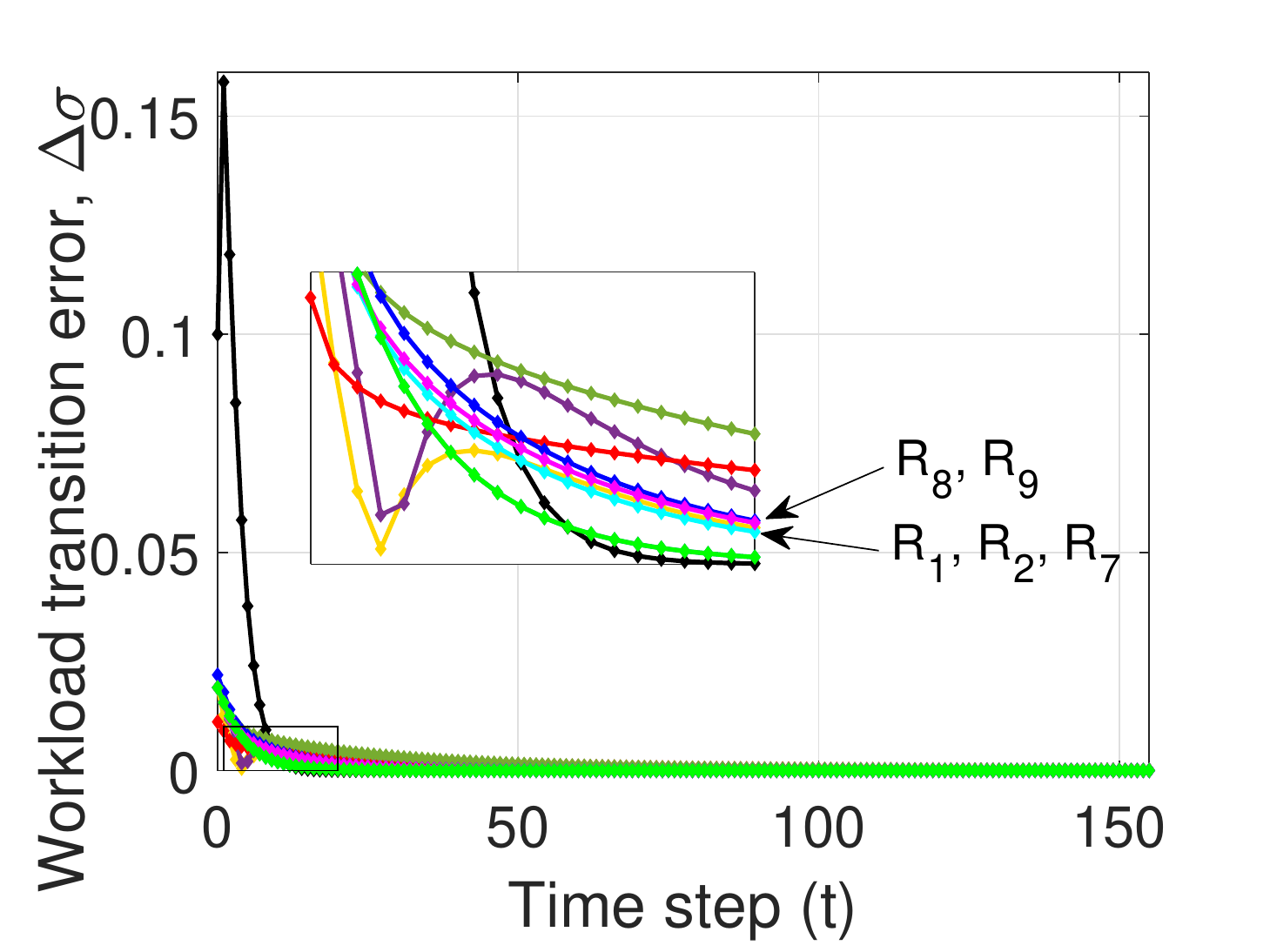}
          \caption{Allocated workload convergence error over time for $K=5$.}
          \label{fig:c32_allocError}
        \end{subfigure}
    \end{minipage}
    \vspace{-4mm}
    \caption{S4: Transition analysis for $m=10$ stationary robots closer to their left region boundary along the horizontal axis, with equal initial workload allocation. $R_3$ is simulated to completely fail with $c^r_3=0$, along with human and robot conditions $c^o_3=0.8$, $c^o_5=0.8$, $c^o_8=0.75$ from $t=0$, and the system adaptively converges to the new workload depending on $q_f$. Green dots represent working robots and red dots represent failed robots. Zoomed sections of plots shown in insets.}
    \label{fig:scalable_2}
    \vspace{-6mm}
\end{figure}

The workload transitioning module of the proposed MH-MR workload allocation framework is a function of $q_f$. We present the effects of different $q_f$ on workload transition with simulation results of $m=10$ robots stationary at all times. Scenario S3 simulates deteriorating human and robot conditions with all robots initially placed at the center of their regions along the horizontal axis; scenario S4 simulates failed robot cases with all robots initially placed closer to the left boundary of their rectangular regions. Odd-indexed robots are assumed to be human-operated while even-indexed robots are assumed autonomous. At $t=0$, the agent conditions are set to $c^o_3=0.8$, $c^o_5=0.8$ and $c^r_8=0.75$ in both scenarios; $R_3$ is set to $c^r_3=0.6$ as deteriorated condition in S3 and $c^r_3=0$ as failed condition in S4 with the rest of the agent conditions remaining at $1$ at all times. Fig. \ref{fig:c31_area} and \ref{fig:c32_area} illustrate the initial setup for S3 and S4 respectively. 

The workload convergence and convergence error plots for S3 shown in Fig. \ref{fig:c31_allocConv} and \ref{fig:c31_allocError}, present a uniform workload transition for all robots to their re-allocated workloads consistent with the setup having all robots initially placed at the center of their regions along the horizontal axis. $R_3$ converged to the lowest allocated workload followed by $R_8$ and $R_5$, while the other robots compensated with increased allocated workload. As such, the workload convergence rate was highest for $R_3$ followed by $R_8$ and $R_5$ with increasingly slower rates respectively following smaller $\Delta \sigma$. The rest of the robots showed the smallest rate of convergence to increased allocated workload with small and equal change in $\Delta \sigma$.

In contrast, S4 where $R_3$ is simulated to fail completely, converges to the zero allocated workload much faster given the larger $\Delta\sigma$ as shown in Fig. \ref{fig:c32_allocConv}. The actual workload convergence rate for $R_3$ was followed by $R_8$ and $R_5$ with increasingly slower rates respectively following smaller $\Delta \sigma$ in comparison. $R_2$ and $R_4$ are both observed to gain a higher workload initially at around $t=15$ due to their close proximity to the largest changing workload allocation in the system for $c^r_3=0$, before reaching an equilibrium workload with the other robots. Between $0<t<25$ with $\sigma_3$ shrinking to zero faster than the other robots, $R_2$ and $R_4$ compensate with a larger share of actual workload temporarily experiencing a faster convergence rate compared to the other agents before the adjustments are propagated to the rest of the robots reaching an equilibrium in the group. The convergence rate of workload allocation for $R_2$ is initially observed slightly higher than $R_4$ for $t<10$ consistent with the proposed workload transition model with $R_2$ initially placed further away from the changing boundary with $R_3$. $R_4$ then shows a higher workload convergence rate between $10<t<15$ as the changing boundary moves away from $R_4$ and closer to $R_2$. The rest of the robots in the group reach an equilibrium workload fairly slowly in comparison, as $R_2$ and $R_4$ adjust over time. 
\begin{figure}[t!]
    \centering
    \includegraphics[width=\linewidth,height=2in]{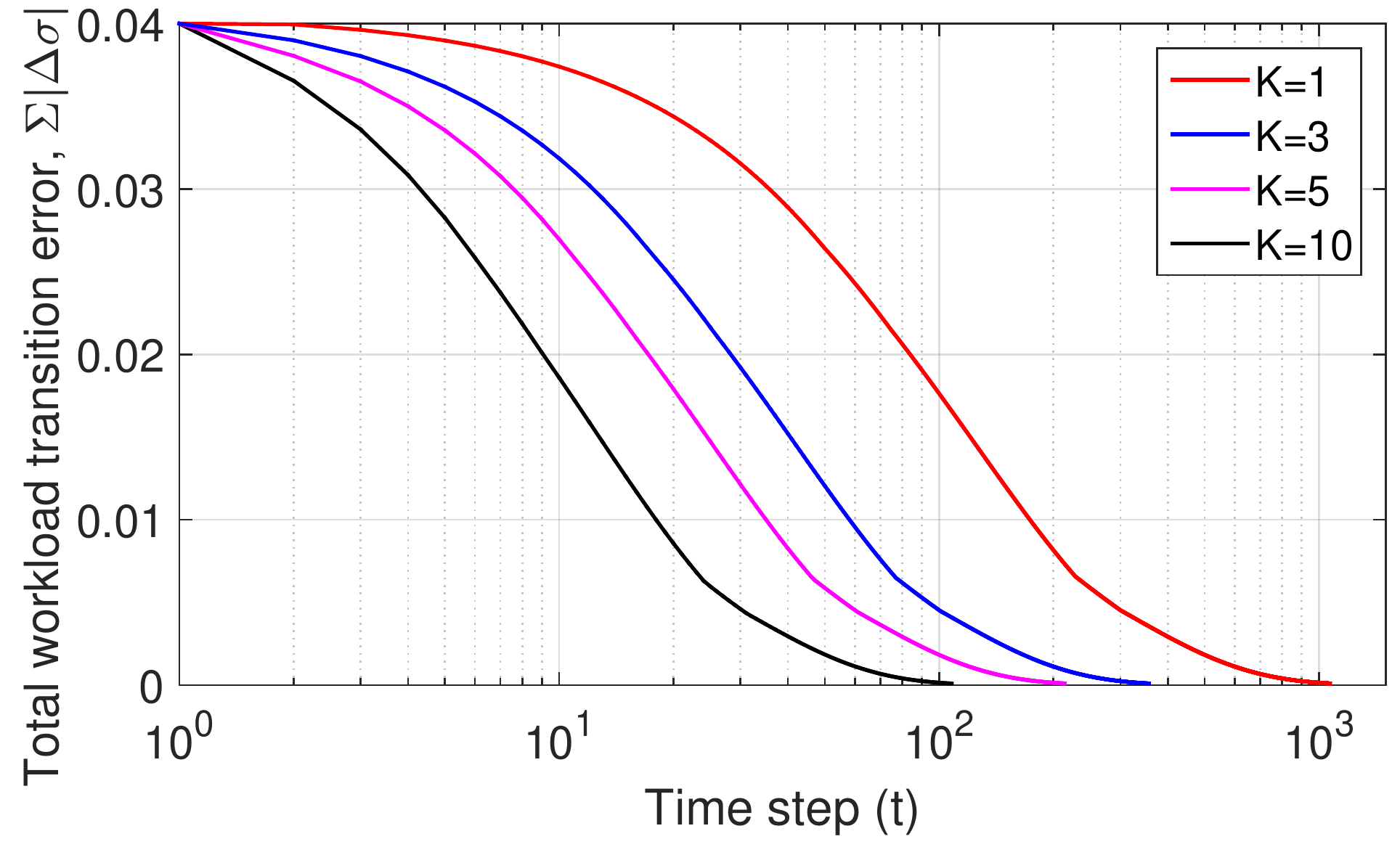}
    \caption{Effect of $K$ on the total workload transition time for $m=50$ robots following scenario S3: human and robot conditions set to $c^o_3=0.8$, $c^o_5=0.8$, $c^r_8=0.75$ and $c^r_3=0.6$. The system adaptively converges to the new workload fastest for $K=10$ and increasingly slower with lower $K$ as expected.}
    \label{fig:diff_K}
    \vspace{-4mm}
\end{figure}

In Fig. \ref{fig:c32_allocError}, the workload transition error of $R_3$ shows an initial error of $0.1$ due to the condition $c^r_3=0$ at $t=0$. Over the next few time steps, the error is observed to rise sharply a little over $0.15$ as the effect of the deteriorated conditions of $R_5$ and $R_8$ are propagated to the transitioning workloads of the rest of the robots including $R_3$; i.e. the error for $R_3$ was compounded with the compensating errors of $R_5$ and $R_8$ before reaching a transitional error of zero. With a complete robot failure in S4, the amount of workload to be re-allocated was larger while considering transitional effects on all agents; hence the workload convergence time for S4 was recorded higher than S3. The workload convergence and the convergence error plot for S4 are shown in Fig. \ref{fig:c32_allocConv} and \ref{fig:c32_allocError} respectively.

\begin{figure}[t]
    \centering
    \includegraphics[width=\linewidth,height=2in]{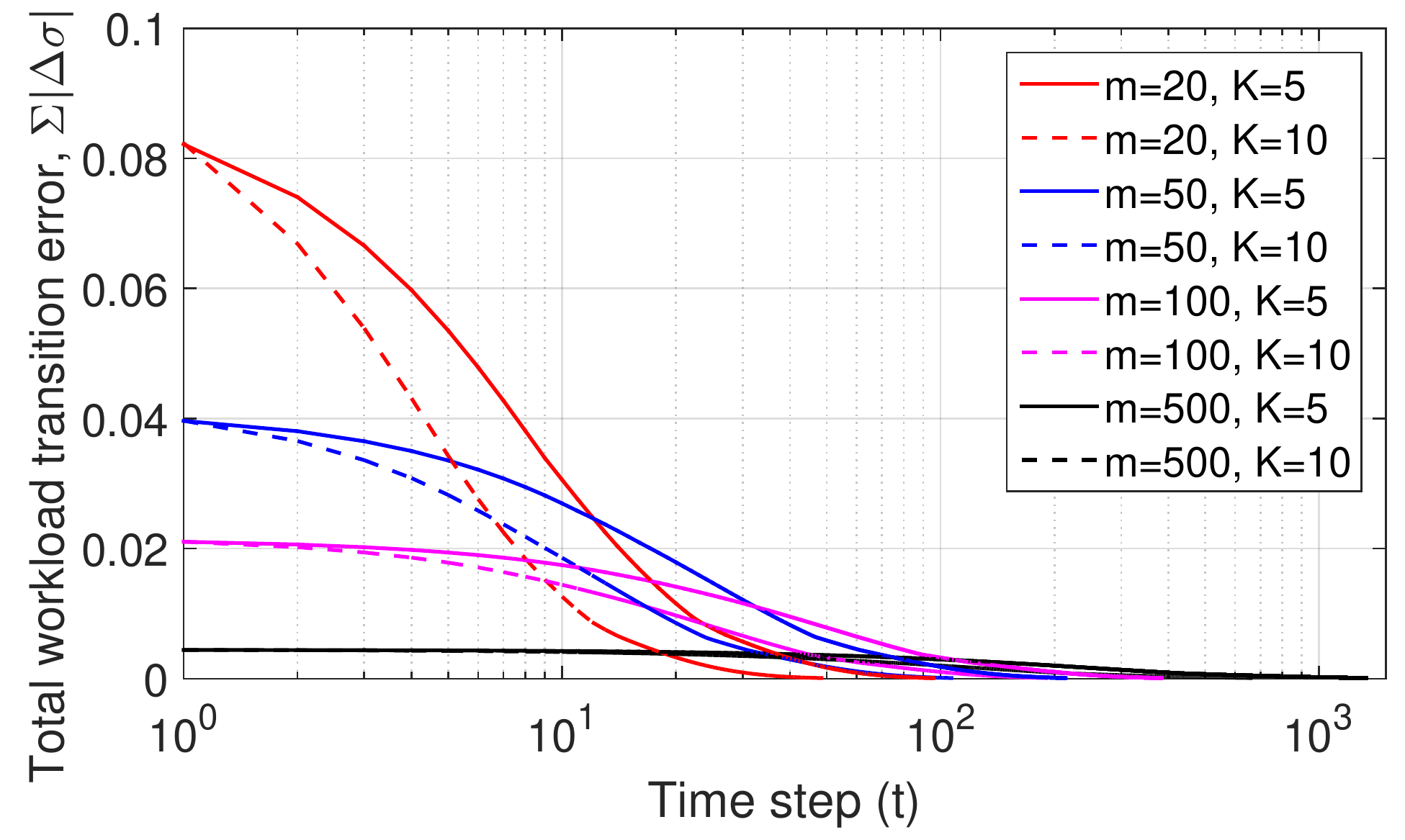}
    \caption{Scalability analysis with $m=20$, $m=50$, $m=100$ and $m=500$ robots following scenario S3: human and robot conditions set to $c^o_3=0.8$, $c^o_5=0.8$, $c^r_8=0.75$ and $c^r_3=0.6$. Effect of $K$ is consistent with larger $K$ yielding faster total workload transition time. Increasingly larger $m$ resulted in increasingly smaller total workload allocation error and longer transition times as expected.}
    \label{fig:diff_m}
    \vspace{-5mm}
\end{figure}

As scalability analysis of the proposed MH-MR workload allocation and transition framework, scenario S3 was repeated for $m=50$ with varying $K$. Fig. \ref{fig:diff_K} plots the total workload transition error along a logarithmic time scale for $K=1$, $K=3$, $K=5$ and $K=10$. The total workload transition error for all cases of $K$ reach zero in finite time. $K=10$ yielded the fastest convergence of the error to zero with increasingly slower rates for lower values of $K$ as expected. Similar observations are made for $m=20$, $m=100$ and $m=500$ robot cases each with $K=5$ and $K=10$ as shown in Fig. \ref{fig:diff_m}; a larger $K$ yielded a faster convergence of the total workload convergence error to zero. The effect of larger $K$ gets smaller with larger $m$; a minor difference is observed for the two $K$ cases for $m=500$. The total initial error was higher for smaller $m$ due to the initially equal distribution of workload assumption of the scenario.

\section{Discussion}\label{sec:dis}
The workload-allocation and transition problem is addressed from a high-level abstraction to maintain generality of its application. The proposed MH-MR framework is suitable for homogeneous and heterogeneous robots (ground, aerial etc.) on homogeneous tasks, given that all robots in the group are capable of completing the homogeneous task of the system independently, and each robot is equipped with all appropriate and relevant low-level controllers. The system is robust to addition and removal (varying $m$) of autonomous and teleoperated robots alike at any time during the mission since each update cycle of the workload allocation process is independent of the previous; the system will simply reallocate the workload accordingly in the next update cycle following Eq. (\ref{eq:S_set}). Workload transition considerations of added robots in the next update cycle are also made with $\sigma(t+1)$ determined with $\sigma(t)=0$ following Eq. (\ref{eq:sigma}). 

The system allows autonomous robots and any number of human operators to teleoperate any number of robots: each human operator may teleoperate multiple robots and multiple human operators may teleoperate a single robot. The graph structure represents the variable human-robot connectivity of the system. The graph connectivity must therefore be updated within the system update cycle $\tau$. Any discontinuity or disconnectedness in the teleoperation or communication graph structure defined in Section \ref{sec:prob_stat} is treated as a failed robot with zero \textit{health}. 

Our current work limits the update cycle to the slowest frequency of the individual modules. However, we acknowledge that it may be improved by considering the highest frequency of all the individual modules and relying on current estimates for the slower modules; an implementation of the Kalman filter for the slower modules may also be used for better current estimates. We identify this potential improvement in the system update cycle as future work on our proposed framework.

We acknowledge that if a large number of agents suffer from deteriorated conditions at once and the rest of the agents are asked to compensate, since the system is designed to ensure that the entire workspace is allocated at all times, it may overwhelm them as well and in turn affect their \textit{health}/\textit{ability} and performance as well. The current framework is unable to consider how much of the total workload can actually be allocated to the given number of agents such that certain agents are not overwhelmed even if their \textit{health}/\textit{ability} are optimal. We identify this as a limitation to the proposed framework and hope to address this issue in our future work. The current framework is therefore only applicable assuming enough agents are in optimal conditions in terms of \textit{health}/\textit{ability}, such that the total workspace could be covered at all times without affecting compensating agents.

The current MH-MR framework assigns workload relying only on agent condition and performance. However, human operators in the system may have different levels of skill, experience and responsiveness despite the measured human condition metric. As such, we acknowledge that with the current design for workload allocation the full potential of the human robot team may not be utilized. Different human operators may also have different working capabilities even under stress or different emotional states which have not been considered in the current system. A number of other complexities also exist on measuring human operator health and condition in the real world in terms of applicability of sensors, the variable calibration requirements and environmental effects that contribute to the huge variety in recorded human behavior \cite{khusainov2013real}. Therefore, as future work of our MH-MR workload allocation framework, we intend to investigate independent human condition assessment and incorporate further human operator attributes in the workload allocation process.

The proposed framework allows multi-human and multi-robots to work together in a given application; robots are free to work autonomously and may also be teleoperated by human operators all the while ensuring that the total work always sums to unity. Therefore, autonomy of the system on the application is shared amongst all individual agents. With the lap time comparison for scenario S1, and failed robot case in scenario S2 presented in the validation section of the manuscript, we established the effectiveness of our proposed workload allocation framework. Therefore, we believe that the proposed MH-MR workload allocation framework is an effective shared-autonomy tool. S3 and S4 presented the effects of $q_f$ on workload allocation followed by the scalability of the system established for various $K$.

The proposed workload allocation and transition function modules are designed to reflect the condition and performance of the autonomous robots, and the humans and robots in teleoperated robots focusing on the working ability of each individual rather than overall optimal system performance. This approach is important for operators (any agents in the system) to believe that the system will consider any deterioration in their \textit{health}/\textit{ability} to work and adjust their workload accordingly, such that they are never overwhelmed. We believe that this function could potentially instill trust in operators (any agents in the system) on the shared-autonomy of the system while working to ensure that they are never overwhelmed with their currently allocated workload.

\section{Conclusion}\label{sec:con}
An adaptive multi-human multi-robot system framework has been proposed that performs real-time workload allocation based on both human operator and robot conditions and on performance, with workload transitional considerations. The design allows compatibility with previously established quantitative human and robot \textit{health} assessments tests; assuming human and/or robot conditions and/or performance are measured with enough accuracy, the modular design of the framework can be used for a wide variety of multi-agent applications, including search and rescue, exploration, surveillance and monitoring based on specific requirements. The system functions independent of the number of humans or robots and is therefore scalable to hosting any number of agents.

The applicability, effectiveness and scalability of the proposed framework was validated experimentally with a MH-MR patrolling application, demonstrating system adaptation to maintain performance despite simulated temporary and permanent, deteriorating human and robot conditions, including complete robot failures. Further work on incorporating modular function blocks on human experience, skill, responsiveness and safety protocols within the work allocation module in the presence of sub-nominal human and/or robot conditions is currently underway, along with field deployment studies of the proposed framework.

\ifCLASSOPTIONcaptionsoff
  \newpage
\fi

\bibliographystyle{IEEEtran}

\bibliography{references}

\begin{IEEEbiography}[{\includegraphics[width=1in,height=1.25in,clip,keepaspectratio]{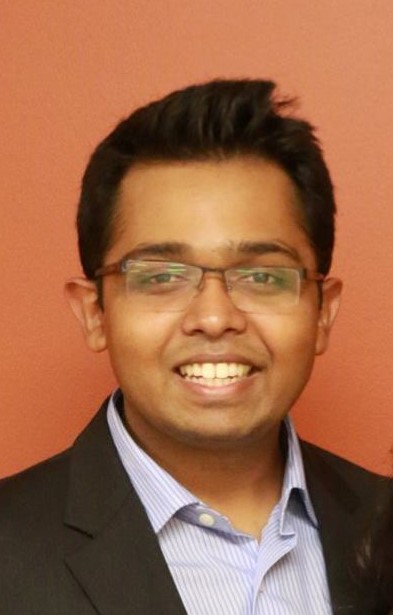}}]{Tamzidul Mina} received the B.S degree in Mechanical Engineering from Purdue University, West Lafayette, USA, in 2012. He is currently pursuing a Ph.D. degree in Mechanical Engineering at Purdue University, West Lafayette, IN.

His research interests include multi-robot systems, control system design for bio-inspired robotic systems with a focus on applications in social group behavior in robotic swarms. 
\end{IEEEbiography}
\vskip 0pt plus -1fil

\begin{IEEEbiography}[{\includegraphics[width=1in,height=1.25in,clip,keepaspectratio]{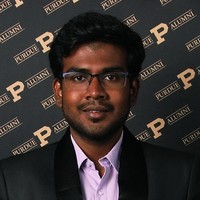}}]{Shyam Sundar Kannan} received the B.E degree in Computer Science and Engineering from Anna University, Chennai, India, in 2016 and the M.S. degree in Computer and Information Technology from Purdue University, West Lafayette, IN, USA, in 2019. He is currently pursuing the Ph.D. degree in Technology at Purdue University, West Lafayette, IN.

From 2016 to 2017, he worked as a Research Assistant at Advanced Geometric Computing Lab, IIT-Madras, Chennai, India. His research interests include SLAM, localization and path planning for multi-agent systems and computational geometry. 
\end{IEEEbiography}
\vskip 0pt plus -1fil

\begin{IEEEbiography}[{\includegraphics[width=1in,height=1.25in,clip,keepaspectratio]{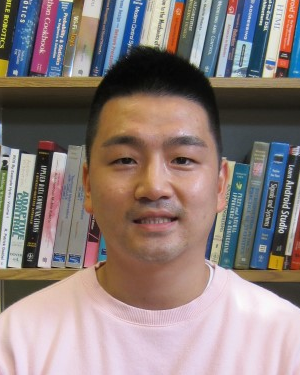}}]{Wonse Jo} received the B.S. in robotics engineering from Hoseo University, South Korea in 2013 and M.S. degrees in electronic engineering from 
the Kyung-Hee University, South Korea, in 2015. He is currently pursuing the Ph.D. degree in computer and information technology at Purdue University, West Lafayette, IN, USA.

His research interests include human-robot interaction, environmental robotics, and assistive robotics. 
\end{IEEEbiography}
\vskip 0pt plus -1fil

\begin{IEEEbiography}[{\includegraphics[width=1in,height=1.25in,clip,keepaspectratio]{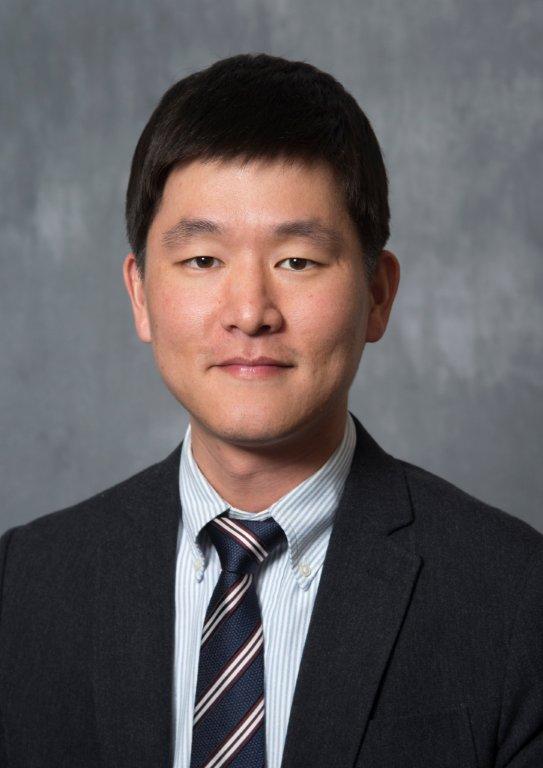}}]{Byung-Cheol Min} (M'14) received the B.S. degree in electronics engineering and the M.S. degree in electronics and radio engineering from Kyung Hee University, Yongin, South Korea, in 2008 and 2010, respectively, and the Ph.D. degree in technology with a specialization in robotics from Purdue University, West Lafayette, IN, USA, in 2014. 

He is an Assistant Professor of Department of Computer and Information Technology and the Director of the SMART Laboratory with Purdue University, West Lafayette, IN, USA. Prior to this position, he was a Postdoctoral Fellow with the Robotics Institute, Carnegie Mellon University, Pittsburgh, PA, USA. His research interests include multi-robot systems, human-robot interaction, robot design and control, with applications in field robotics and assistive technology and robotics.

He is a recipient of the NSF CAREER Award in 2019, Purdue PPI Outstanding Faculty in Discovery Award in 2019, and Purdue CIT Outstanding Graduate Mentor Award in 2019.
\end{IEEEbiography}


\end{document}